\documentclass[review,3p]{elsarticle}

\usepackage{verbatim}
\usepackage{lineno,hyperref,graphicx}
\usepackage{morefloats}
\usepackage[usenames, dvipsnames]{color}
\usepackage{soul}
\usepackage{changes}
\usepackage{perpage}
\usepackage[hang,tight]{subfigure}
\MakePerPage{footnote}
\journal{Nuclear Instruments and Methods in Physics Research Section A}
\begin{document}

\begin{frontmatter}

\title{Performance of a Large-area GEM Detector Read Out with\\ Wide Radial Zigzag Strips}
\tnotetext[mytitlenote]{This draft has been submitted to NIMA for review.}
%\tnotetext[mytitlenote]
\author{Aiwu Zhang\corref{cor1}}
\cortext[cor1]{Corresponding author: Tel. +1(321)674-7339; Email \href{mailto:azhang@fit.edu}{\textbf{azhang@fit.edu}}.}

\author{Vallary Bhopatkar\corref{}}
\author{Eric Hansen}
\author{Marcus Hohlmann}
%\ead{hohlmann@fit.edu}
\author{Shreeya Khanal}
\author{Michael Phipps}
\author{Elizabeth Starling}
\author{Jessie Twigger}
\author{Kimberly Walton}

\address{Department of Physics and Space Sciences, Florida Institute of Technology, Melbourne, FL 32901, USA}




\begin{abstract}
%\begin{linenumbers}
A 1-meter-long trapezoidal Triple-GEM detector with wide readout strips was tested in hadron beams at the Fermilab Test Beam Facility in October 2013. The readout strips have a special zigzag geometry and run along the radial direction with an azimuthal pitch of 1.37~mrad to measure the azimuthal $\phi$-coordinate of incident particles. The zigzag geometry of the readout reduces the required number of electronic channels by a factor of three compared to conventional straight readout strips while preserving good angular resolution. The average crosstalk between zigzag strips is measured to be an acceptable 5.5\%. The detection efficiency of the detector is  (98.4~$\pm$~0.2)\%. When the non-linearity of the zigzag-strip response is corrected with track information, the angular resolution is measured to be (193~$\pm$~3)~$\mu$rad, which corresponds to 14\% of the angular strip pitch. Multiple Coulomb scattering effects are fully taken into account in the data analysis with the help of a stand-alone Geant4 simulation that estimates interpolated track errors.  
%\end{linenumbers}
\end{abstract}

\begin{keyword}
MPGD; GEM; Zigzag readout strip; Beam test; Angular resolution.
%\texttt{elsarticle.cls}\sep \LaTeX\sep Elsevier \sep template
%\MSC[2010] 00-01\sep  99-00
\end{keyword}

\end{frontmatter}

\section{Introduction} \label{Introduction}
Spatial resolution is always an important parameter for position-sensitive particle detectors, such as Micro-Pattern Gaseous Detectors~(MPGD). It is well known that a Gas Electron Multiplier~(GEM)~\cite{Sauli97} detector with straight readout strips of 400~$\mu$m pitch can reach 50-70~$\mu$m resolution \cite{COMPASSGEM, KondoNIMA}. In applications with large-area GEM detectors, the use of readout strips with such small pitch will quickly lead to a large number of required electronic channels and consequently to significant system cost incurred by the readout electronics. Readout structures employing pads or strips with a chevron or zigzag geometry were proposed and studied in the past to address this issue for MPGDs and other gaseous detectors~\cite{zz85, zz89, zz97, zzBYu03, zzBYu05}. A zigzag strip covers a larger area than a standard straight strip, so that zigzag strips can reduce the number of electronic channels needed to read out a given detector area. The challenge then is to make sure that the spatial resolution is still adequate and that the crosstalk between strips is well controlled.
 
We have previously demonstrated for small Triple-GEM detectors that parallel zigzag strips achieve good spatial resolution as the sensitivity of the charge sharing among strips to the hit position is enhanced between the interleaved ``zigs'' and ``zags'' of adjacent strips~\cite{IEEEzigzag}. Here we expand this approach to large-area GEM detectors with a radial readout strip geometry that is appropriate for detector systems with a disk or ring geometry as is commonly used in collider experiments. We report results from a beam test of a 1-meter-long GEM detector read out with wide radial zigzag strips.  

This paper is organized as follows: In section \ref{theDetector} the design of the zigzag readout board is described in detail; in section \ref{setup} the beam test configuration and data acquisition system are introduced; in section \ref{basic_results} the data analysis methods and basic performances of the zigzag GEM detector are discussed; in section \ref{resolution} the angular resolution, i.e.\ the spatial resolution in azimuthal $\phi$-coordinate, of the zigzag GEM detector is studied under different operating conditions.

\section{The GEM detector and the zigzag readout strips} \label{theDetector}
\begin{figure}[h]
 \centering
  \includegraphics[width=0.7\textwidth]{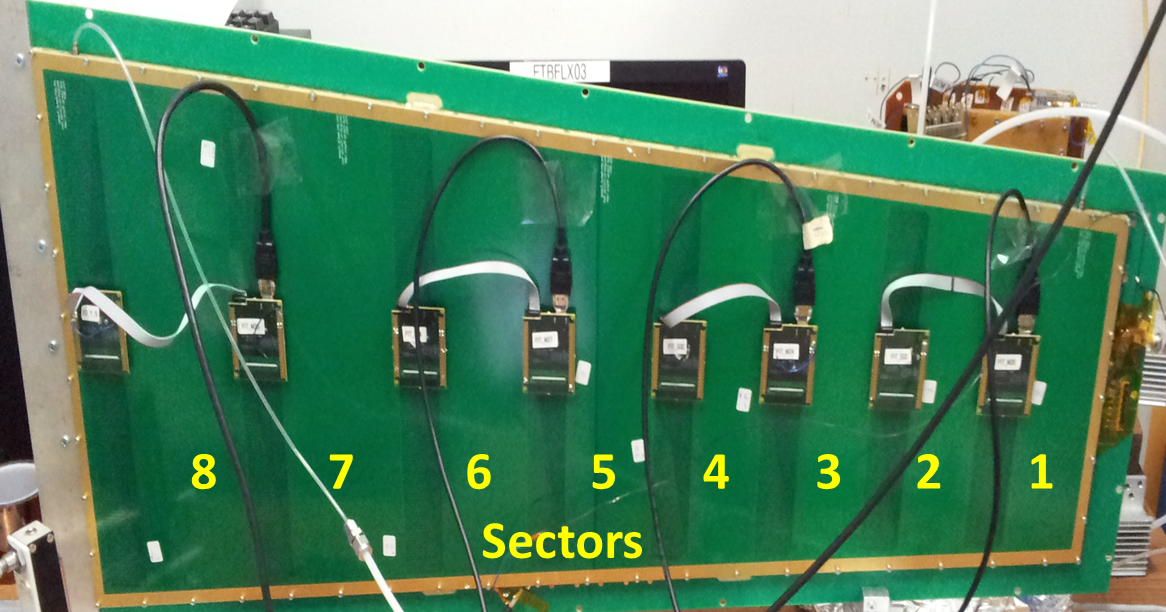}
  \vspace{-0.5cm}
  \caption{The 1-meter-long trapezoidal GEM detector with zigzag readout strips. The detector has eight sectors; each comprises 128 radial zigzag strips and is read out with an APV25 hybrid board.}
  \label{zigzagGEM}
\end{figure}

The 1-meter-long Triple-GEM detector has the shape of a trapezoid with 22~cm width at the narrow end and 45~cm width at the wide end (Fig.~\ref{zigzagGEM}). It is based on a prototype for the CMS muon GEM upgrade at LHC~\cite{CMSGEM, CMSGEMTDR}, but has a modified readout board. The gas gaps (drift, transfer 1, transfer 2, induction) in this GEM detector are 3/1/2/1~mm.

\begin{figure}[h]
  \centering
  \includegraphics[width=10cm]{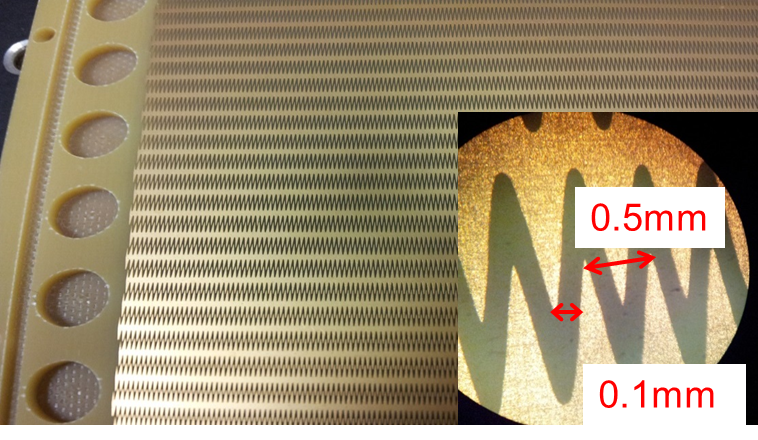}
  \vspace{-0.5cm}
  \caption{A photo of the side of the readout pcb that features the radial zigzag strips. The inset is an image of the zigzag structure taken with a microscope. The strips run in radial direction with an angular pitch of 1.37~mrad. The distance between two tips in neighboring strips is 0.1~mm, and the distance between two tips in the same strip is 0.5~mm.}
  \label{zigzagDetail}
\end{figure}

On the printed circuit board (pcb) used for the readout, the zigzag strips are radially divided into eight sectors with lengths 10~cm (15~cm) at the narrow (wide) end of the trapezoid and with 128 strips per sector (Fig.~\ref{zigzagGEM}). In a collider geometry these sectors would correspond to different ranges of pseudorapidity $\eta$. From the narrow end to the wide end, the sectors are labeled from one to eight. The reason for this partitioning is to provide some coarse information on the radial position of hits and to have readout strips with reasonably low capacitance. In Fig.~\ref{zigzagDetail}, the details of the zigzag structure are shown. The zigzag strips run in radial direction and measure the azimuthal $\phi$ coordinate. The opening angle between the first strip and the last strip in a sector is 10$^\circ$ and the angular pitch between two neighboring strips is 1.37~mrad. The spacing between zigzag tips on adjacent strips is 0.1~mm and the distance between two tips in one strip is 0.5~mm. These geometric parameters are the same for all strips in all eight sectors on the readout board; consequently the strip width increases from $\sim$2.5~mm at the narrow end to $\sim$4.5~mm at the wide end. The total number of zigzag strips on the readout board is 1,024, to be compared with, e.g., 3,072 straight strips in the standard CMS GEM detector that the zigzag detector is derived from~\cite{CMSGEMTDR}. This means that 2/3 of the electronics can be eliminated by using zigzag strips. 

A miniaturized ceramic high voltage divider is used for powering the GEM detector. The HV divider has a resistor chain with a total resistance of 4.4 M$\Omega$ and connects to each electrode of the GEM detector through eight pads on the drift board. One HV input channel is connected to the drift electrode and potentials between any two neighboring detector electrodes are provided through resistors in the HV divider. More details of the HV divider can be found in reference~\cite{CMSGEMTDR}.

\section{Beam test configuration and data acquisition system}\label{setup}
\begin{figure}[h]
  \begin{center}
     \subfigure{
          \label{fig3:top}%%
          \begin{minipage}[b]{0.5\textwidth}
              \centering
              \includegraphics[width=8cm,height=6cm]{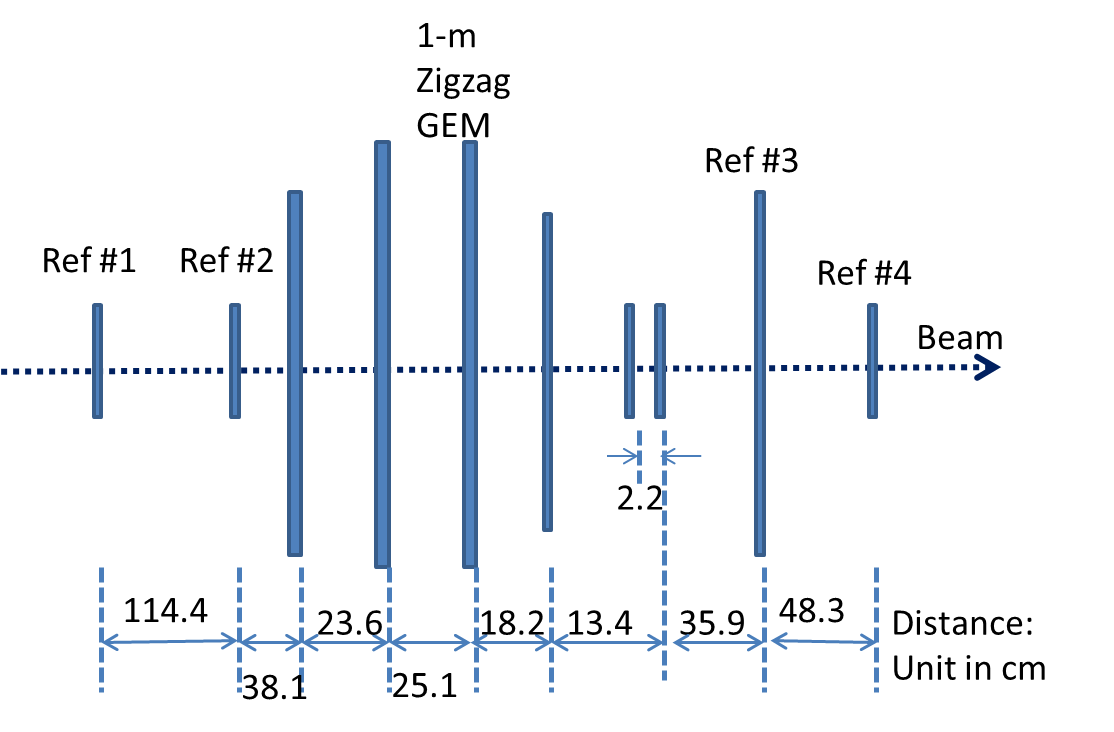}
          \end{minipage}}\newline%
    \subfigure{
          \label{fig3:bottom}%%
          \begin{minipage}[b]{0.5\textwidth}
              \centering
              \includegraphics[width=8cm,height=6cm]{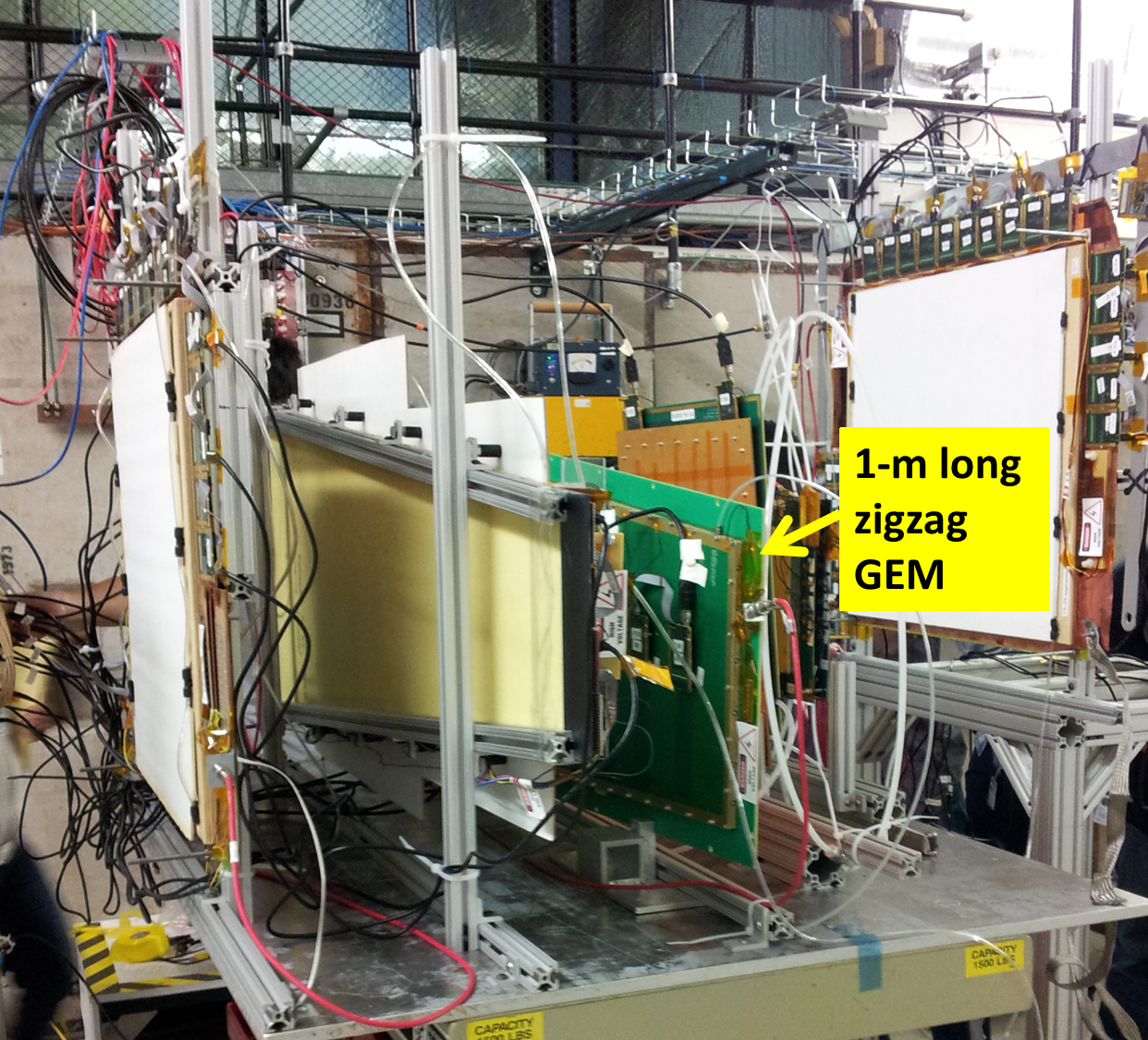}
          \end{minipage}}\newline%
    \vspace{-1cm}
    \caption{Top: The overall configuration of all tracking GEM detectors in the beam line (not to scale). Bottom: A photo of the zigzag GEM detector (pointed out by the yellow arrow) in the beam line.}
    \label{detectorConfig}
  \end{center}
\end{figure}

The zigzag GEM detector was tested as a tracking detector at the Fermilab Test Beam Facility~(FTBF) in October 2013. The eRD6-FLYSUB consortium~\footnote{eRD6-FLYSUB is a consortium of researchers from \textbf{FL}orida Institute of Technology, \textbf{Y}ale University, \textbf{S}tony Brook University, \textbf{U}niversity of Virginia, \textbf{B}rookhaven National Lab, formed to carry out R\&D of tracking and particle identification detectors for a future electron-ion collider.} conducted this beam test with a variety of GEM detectors. Fig.~\ref{detectorConfig}~(top) shows a diagram of the detector configurations in the tracking system, where a set of four GEM detectors with 2-D readout acts as a reference tracker (noted as ``Ref'' in the figure). The zigzag GEM detector was installed on a movable table in the center between the tracker detectors (Fig.~\ref{detectorConfig}, bottom). All detectors were operated with an Ar/CO$_2$ 70:30 gas mixture during the entire test period.

\begin{figure}[h]
  \begin{center}
     \subfigure{
          \label{fig4:top}%%
          \begin{minipage}[b]{0.5\textwidth}
              \centering
              \includegraphics[width=8cm,height=6cm]{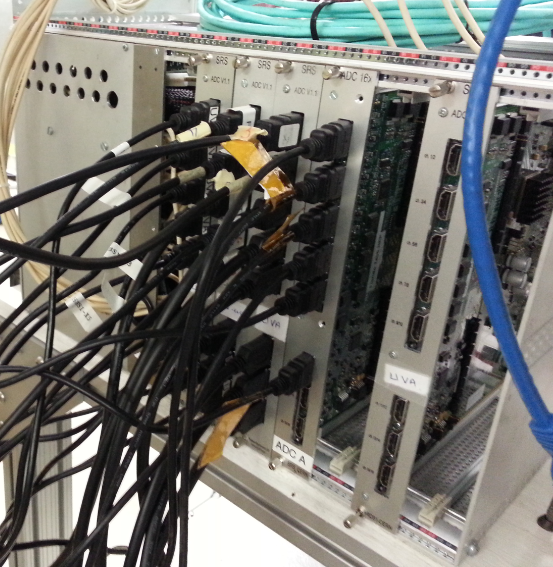}
          \end{minipage}}\newline%
    \subfigure{
          \label{fig4:bottom}%%
          \begin{minipage}[b]{0.5\textwidth}
              \centering
              \includegraphics[width=8cm,height=6cm]{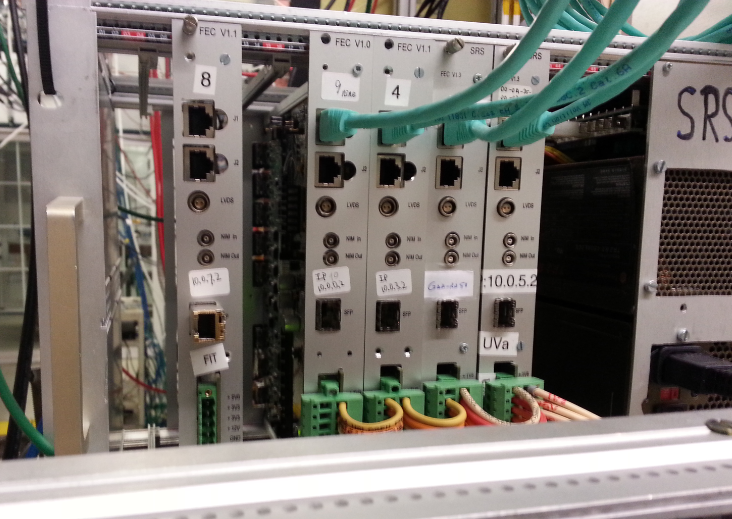}
          \end{minipage}}\newline%
    \vspace{-1cm}
    \caption{The Scalable Readout System used in the beam test. Top: HDMI cables from APV25 hybrids connect to ADCs that are connected to Front-End Concentrators (FECs). Bottom: The FECs in turn communicate with a DAQ PC through Gigabit ethernet. One ADC+FEC unit reads out 16 APV25 hybrids, which corresponds to 2,048 channels.}
    \label{fig4}
  \end{center}
\end{figure}

The RD51 Scalable Readout System (SRS)~\cite{SRS} was used to read out all ten detectors in the beam test. With 64 frontend hybrids carrying APV25 chips~\cite{APVUserGuide} and four SRS Frontend-Concentrator/ADC combinations, 8,192 channels were read out simultaneously~(Fig.~\ref{fig4}). Strip signals were digitized at 40 MHz. Readout was triggered by coincidence signals from plastic scintillators placed in the beam line. Data were transferred via Gigabit ethernet to a PC that acquired them using the DATE software and monitored them online in the \textbf{A}utomatic \textbf{MO}nito\textbf{R}ing \textbf{E}nvironment~(AMORE) (both software suites were originally developed for the ALICE experiment and later adapted by the RD51 collaboration for the SRS)~\cite{ALICEGuide}.

\section{Data analysis and offline results} \label{basic_results}
The data are analyzed with the AMORE software package as well; raw events are decoded and basic information on the events are retrieved. Events with multiple hits in any detector are excluded in the analysis for simplicity. Beam profiles are checked first to characterize the impinging beams. Then basic performance characteristics of the zigzag GEM detector are analyzed, such as strip multiplicity of strip clusters, i.e.\ the number of strips with induced charge on the strips above a certain threshold, strip-cluster charge distribution and detection efficiency. For simplicity, we will refer to a strip cluster as a ``cluster" throughout the remainder of this paper. Crosstalk between zigzag strips is also investigated in this section.

\subsection{Beam profiles}
\begin{figure}[!htb]
  \begin{center}
        \vspace{-1cm}
        \subfigure[20GeV/c mixed beam]{
          \label{fig5:20GeV}%%
          \begin{minipage}[b]{0.5\textwidth}
            \centering
            \includegraphics[width=7cm]{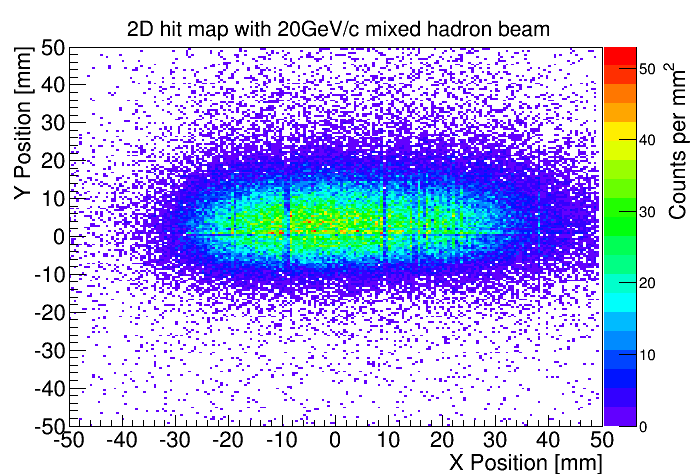}
          \end{minipage}}%
      % \vspace{-2cm}
        \subfigure[25GeV/c mixed beam]{
          \label{fig5:25GeV}%%
          \begin{minipage}[b]{0.5\textwidth}
            \centering
            \includegraphics[width=7cm]{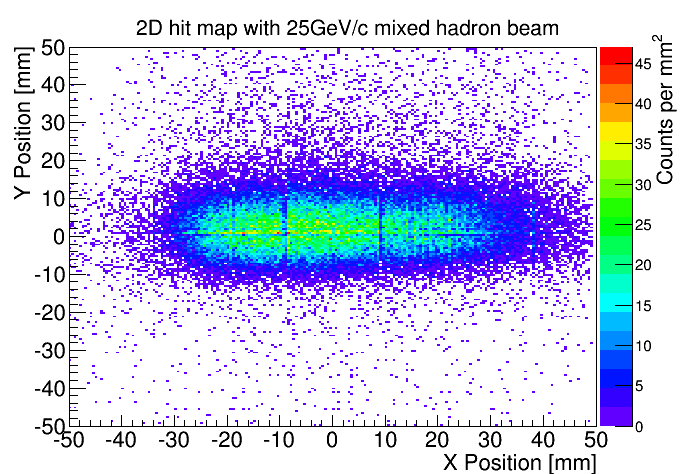}
          \end{minipage}}\newline%
       % \vspace{-2cm}
        \subfigure[32GeV/c mixed beam]{
          \label{fig5:32GeV}%%
          \begin{minipage}[b]{0.5\textwidth}
            \centering
            \includegraphics[width=7cm]{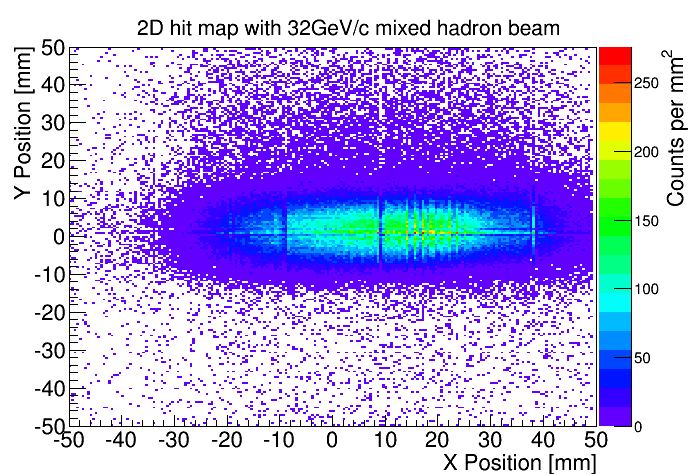}
          \end{minipage}}%
      % \vspace{-2cm}
        \subfigure[120GeV/c proton beam]{
          \label{fig5:120GeV}%%
          \begin{minipage}[b]{0.5\textwidth}
            \centering
            \includegraphics[width=7cm]{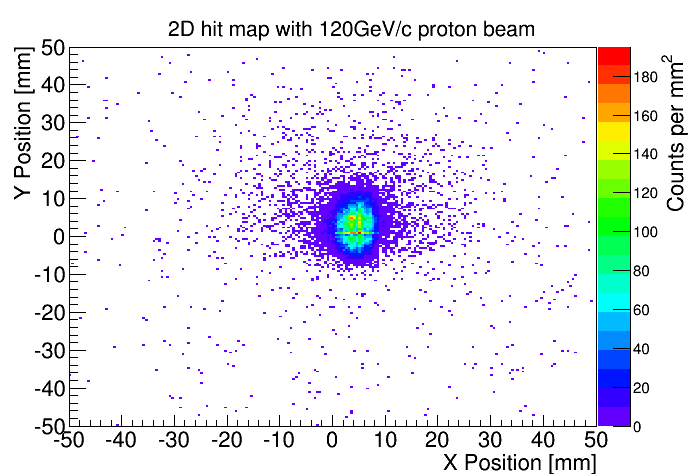}
          \end{minipage}}%%
        \vspace{-0.3cm}
        \caption{2D maps of hits recorded in the first tracker detector for different beam momenta. Only events with hits in all four trackers are shown. As expected, the beam spot of the primary proton beam is much smaller than the beam spot of the secondary mixed hadron beams. The horizontal and vertical lines visible on these plots are due to grid spacers in the third tracker detector that locally reduce the detection efficiency.}
        \label{fig5}
  \end{center}
\end{figure}

During the beam test, we tested detectors with the primary 120~GeV/c proton beam and with secondary beams containing mixed hadrons~(mainly pions with an admixture of kaons and protons) at 20, 25, and 32~GeV/c momenta. Most of the data for the zigzag GEM detector were taken with mixed hadron beams. The 2D hit maps of the tracker detectors~(Fig.~\ref{fig5}) show that the beam spot of the pure proton beam is $\sim$2~cm in diameter, while in the case of mixed hadron beams it is $\sim$7~cm in X direction (horizontal) and $\sim$3~cm in Y direction (vertical).

\subsection{Basic performance characteristics of the zigzag GEM detector}

\subsubsection{Cluster charge vs.\ HV}
Before taking data for any scenario, we took a pedestal run with 5k events at standby detector voltages so that there was no gas gain. To find a cluster in data runs with full voltage, we set a 5$\sigma$ threshold cut where $\sigma$ is the width of the pedestal distribution. If N contiguous strips are fired in a detector with individual strip charges higher than the 5$\sigma$ cut, then we will consider this as a cluster with strip multiplicity N in the detector and accept this event.

\begin{figure}[h]
  \begin{center}
     \subfigure{
          \label{fig6:top}%%
          \begin{minipage}[b]{0.5\textwidth}
              \centering
              \includegraphics[width=8cm,height=6cm]{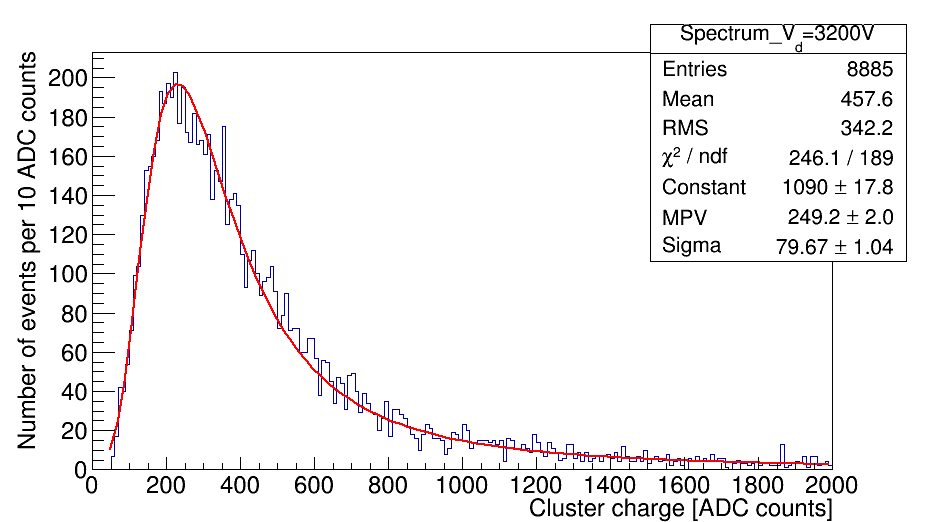}
          \end{minipage}}\newline%
    \subfigure{
          \label{fig6:bottom}%%
          \begin{minipage}[b]{0.5\textwidth}
              \centering
              \includegraphics[width=8cm,height=6cm]{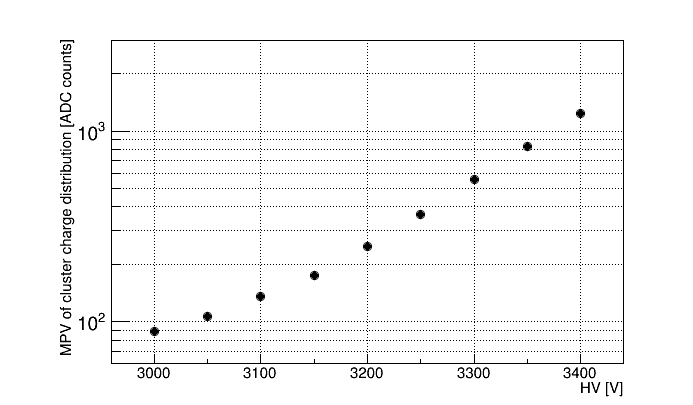}
          \end{minipage}}\newline%
    \vspace{-1cm}
    \caption{Top: Cluster charge distribution measured in central sector 5 of the zigzag GEM detector at $\mathrm{V_{drift}}$ = 3200~V with 25~GeV/c mixed hadron beam and fitted to a Landau distribution. Bottom: Most probable values (MPV) of the measured Landau distributions vs.\ $\mathrm{V_{drift}}$. (1 ADC count corresponds to about 0.0371~fC.)}
    \label{fig6}
  \end{center}
\end{figure}

\begin{figure}[h]
  \centering
  \includegraphics[width=12cm, height=7cm]{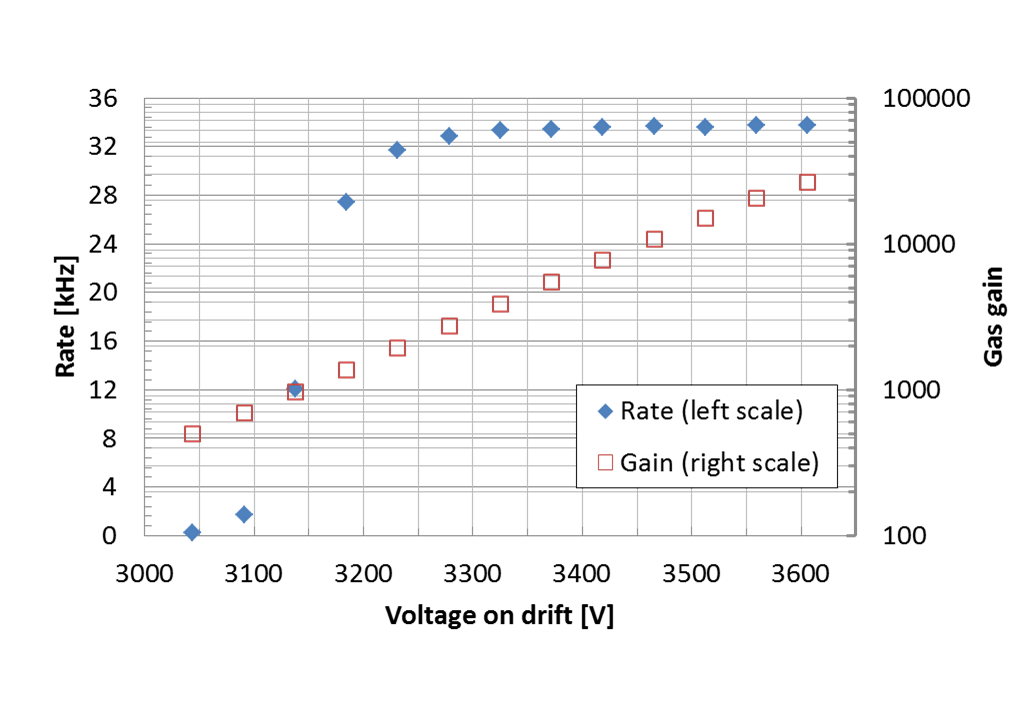}
  \vspace{-0.5cm}
  \caption{Rate and gain vs.\ $\mathrm{V_{drift}}$ measured in central sector 5 of the zigzag GEM detector with Ar/CO$_{2}$ 70:30. As opposed to all other measurements discussed in this paper, this measurement was performed after the beam test at sea level in our lab in Florida with x-rays. The x-rays were generated by an AMPTEK Mini-X x-ray gun with gold anode that was operated at 10~kV accelerating voltage. Errors are smaller than marker sizes.}
  \label{figGain}
\end{figure}

The zizgzag GEM detector was tested at different high voltages $\mathrm{V_{drift}}$ applied to the drift electrode when a 25~GeV/c mixed hadron beam was impinging on central readout sector number five (Fig.~\ref{zigzagGEM}). Fig.~\ref{fig6}~(top) shows the total cluster charge distribution  at $\mathrm{V_{drift}}$ = 3200~V in terms of ADC counts (1 ADC $\simeq$ 0.0371~fC). The measured cluster charge distribution fits well to a Landau function. Fig.~\ref{fig6}~(bottom) shows the most probable values~(MPV) of the Landau fits as a function of $\mathrm{V_{drift}}$; this yields an approximately exponential curve as expected for an MPGD. The gas gain during the beam test is estimated to range from 660~(at $\mathrm{V_{drift}=3000~V}$) to $\mathrm{10^4}$~(at $\mathrm{V_{drift}=3400~V}$) based on the measured mean cluster charge. After the beam test, we measured the gas gain of this detector operated with Ar/CO$_{2}$ 70:30 in our lab in Florida (sea level) with X rays (AMPTEK Mini-X X-ray generator with gold anode, set at 10~kV and 5~$\mu$A) impinging on  central sector five. A good rate plateau is obtained and the gain reaches close to $\mathrm{3\times10^4}$ at $\mathrm{V_{drift}=3600~V}$ (Fig.~\ref{figGain}) under those conditions. 

\subsubsection{Strip multiplicity vs.\ HV}
\begin{figure}[h]
  \begin{center}
     \subfigure{
          \label{fig8:top}%%
          \begin{minipage}[b]{0.5\textwidth}
              \centering
              \includegraphics[width=8cm,height=5cm]{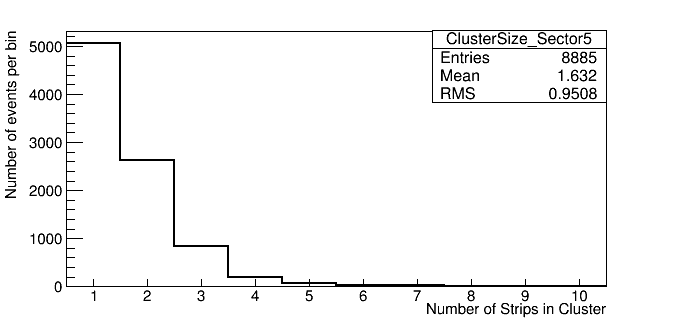}
          \end{minipage}}\newline%
    \subfigure{
          \label{fig8:bottom}%%
          \begin{minipage}[b]{0.5\textwidth}
              \centering
              \includegraphics[width=8cm,height=5cm]{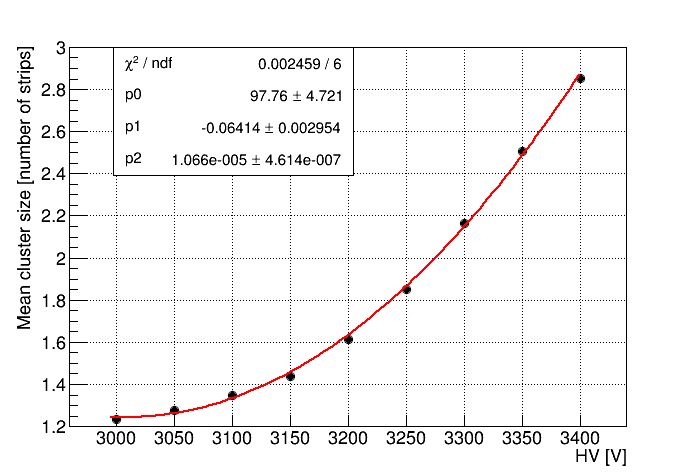}
          \end{minipage}}\newline%
    \vspace{-1cm}
    \caption{Top: Strip multiplicity for clusters in the zigzag GEM detector measured at $\mathrm{V_{drift}= 3200~V}$ in central sector 5 with a 25~GeV/c mixed hadron beam. Bottom: The mean strip multiplicity vs.\ $\mathrm{V_{drift}}$ follows a quadratic function. Errors are smaller than marker sizes. }
    \label{fig8}
  \end{center}
\end{figure}

The strip multiplicity distribution of clusters measured at $\mathrm{V_{drift}}$ = 3200~V in central sector five is shown in Fig.~\ref{fig8}~(top); the mean strip multiplicity is 1.63. This mean value is found to increase quadratically with $\mathrm{V_{drift}}$ (Fig.~\ref{fig8}, bottom), and at the highest tested voltage~(3400~V) its maximum is just below three strips. This is expected since the strip width in that radial region is $\sim$3.6 mm.

\subsubsection{Detection efficiency vs.\ HV}
\begin{figure}[h]
  \centering
  \includegraphics[width=12cm, height=7cm]{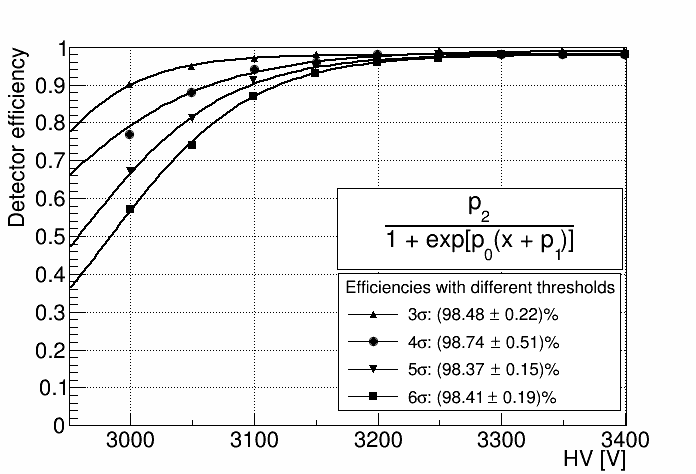}
  \vspace{-0.5cm}
  \caption{Detection efficiency in central sector 5 of the zigzag GEM detector for 25~GeV/c mixed hadron beam as a function of $\mathrm{V_{drift}}$ and for different hit thresholds (errors are smaller than marker size). The applied thresholds are multiples of the width ($\sigma$) of the pedestal distributions for individual strips. The data are fit to sigmoid functions as given in the inset to determine the efficiencies on plateau.}
  \label{fig9}
\end{figure}

The detection efficiency of the zigzag GEM detector for mixed hadrons, which are close to minimally ionizing in this test beam, is obtained from the ratio of the number of observed hits to the total number of triggered events. The measured efficiency is (98.4~$\pm$~0.2)\% on the plateau when a 5$\sigma$ threshold cut is applied to the measured strip charge.  Fig.~\ref{fig9} shows efficiencies with different threshold cuts demonstrating that the plateau efficiency is not affected significantly by the applied threshold.

\subsubsection{Response uniformity from position scan data}
\begin{figure}[h]
  \centering
  \includegraphics[width=12cm, height=7cm]{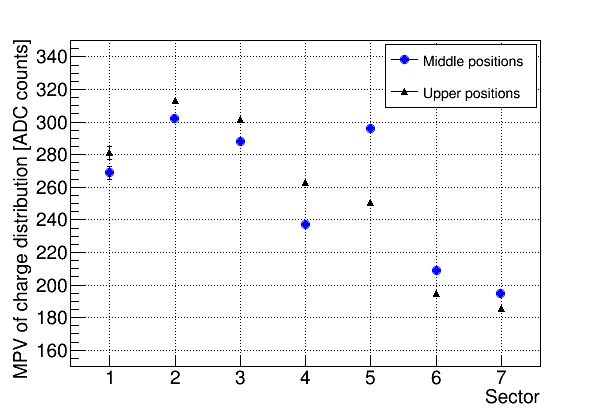}
  \vspace{-0.5cm}
  \caption{Most probable value of cluster charge distributions from Landau fits in different sectors~($\mathrm{V_{drift}}$ = 3200~V). Two points were measured in each sector. ``Middle" means the position is roughly in the center of the sector and $\sim$60~mm lower than the ``Upper" position in that sector. Errors are smaller than marker sizes for higher sector numbers.}
  \label{fig10}
\end{figure}

The zigzag GEM detector was tested (at $\mathrm{V_{drift}}$ = 3200~V) with 20~GeV/c mixed hadron beams impinging on two different spots in each readout sector, approximately 60~mm apart in the vertical direction, so that the response uniformity of the chamber could be examined.  Fig.~\ref{fig10} shows the MPVs of Landau fits to cluster charge distributions measured in sectors one to seven at these different positions. The sector eight was not tested because it could not be reached by moving the stage that the detector was placed on. The response varies by $\sim$25\% and is lower in sectors six and  seven, which are on the wide side of the trapezoidal chamber. This non-uniformity is likely caused by a known slight bending of the drift board that was occurred when the GEM foils were stretched during assembly.

\subsubsection{Crosstalk among zigzag strips}
\begin{figure}[h]
  \begin{center}
     \subfigure{
          \label{fig11:top}%%
          \begin{minipage}[b]{0.5\textwidth}
              \centering
              \includegraphics[width=8cm,height=6cm]{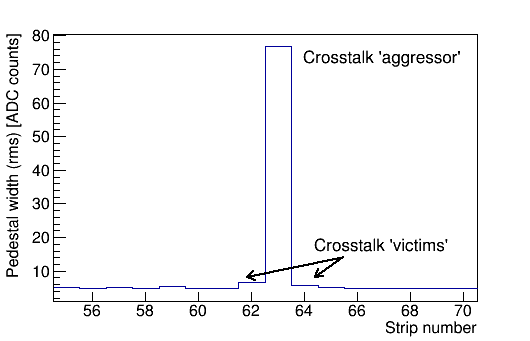}
          \end{minipage}}\newline%
    \subfigure{
          \label{fig11:bottom}%%
          \begin{minipage}[b]{0.5\textwidth}
              \centering
              \includegraphics[width=8cm,height=6cm]{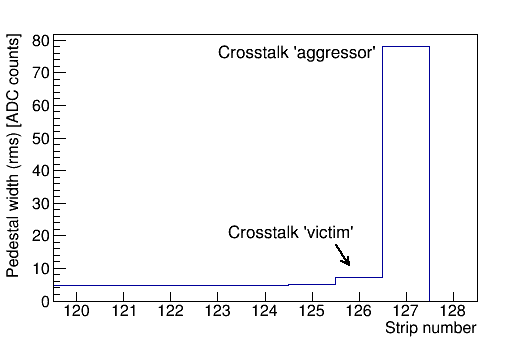}
          \end{minipage}}\newline%
    \vspace{-1cm}
    \caption{Pedestal widths (rms) observed for zigzag strips in the vicinity of two noisy strips (``aggressors") in one sector. Pedestal widths in   
             adjacent strips (``crosstalk victims") are slightly higher than the average pedestal width due to crosstalk from the aggressors.}
    \label{fig11}
  \end{center}
\end{figure}

We were able to investigate the crosstalk between zigzag strips with the beam test data due to a minor mistake that was made in the production procedure of the readout pcb, which accidentally connected strips number 63 and 127 in each sector to the ground plane. The large input capacitance presented by these grounded strips to the amplifier in the APV chip causes large noise as can be seen, for example, in the distribution of the pedestal widths for one of the sectors (Fig.~\ref{fig11}). These plots also show that the adjacent strips 62, 64 and 126 are victims of crosstalk as their pedestal widths are slightly higher than the average across all other strips.

\begin{figure}[h]
  \centering
  \includegraphics[width=10cm]{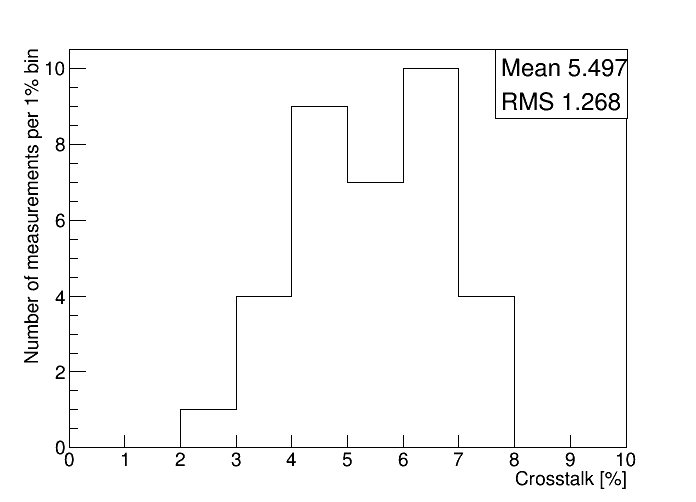}
  \vspace{-0.5cm}
  \caption{Overall crosstalk measured for crosstalk victims in two pedestal runs and in several sectors of zigzag GEM.}
  \label{figOverallCrossTalk}
\end{figure}

We estimate the crosstalk as $\mathrm{\sqrt{rms_{victim}^{2}-rms_{avg}^{2}}/rms_{aggressor}}$, where $\mathrm{rms_{avg}}$ is the average pedestal width observed for strips not affected by crosstalk, $\mathrm{rms_{aggressor}}$ is the pedestal width on the noise source strips (63 or 127), and $\mathrm{rms_{victim}}$ is the pedestal width on a strip adjacent to the noise source strip. We find an average crosstalk among the zigzag strips of ($5.5~\pm~0.2$)\% with an rms width of 1.3\% (Fig.~\ref{figOverallCrossTalk}).

\section{Angular resolution studies}\label{resolution}
\subsection{Multiple Coulomb scattering in detector materials during beam test}
In the beam test, the total radiation length of the detector materials in the setup composed of ten GEM detectors was about 14\% (Table~\ref{mytable}), which impacts the spatial resolution analysis due to multiple Coulomb scattering (MCS) of the tracks in the material. For example, the rms width of the scattering angle distribution due to MCS in that amount of materal is estimated to be 147~$\mu$rad for 32~GeV/c hadrons using the standard MCS formula~\cite{PDG_multipleScattering}. 

\begin{table}[h]
\caption[Table caption text]{Material estimate for the tracking detectors in the FNAL beam test.}
\begin{tabular}{|r||c|c|c|c|}
  \hline
  Detector & Gas gaps & Window mat./     & Readout mat./       & Rad. Len.  \\ 
           &     [mm] &   thickness [mm] & thickness [mm]                & [\%$X_0$] \\ \hline \hline
  Tracker 1 & 3/2/2/2 & Mylar/$\sim$~0.1 & G10/kapton/honeycomb & 0.32 \\ \hline
  Tracker 2 & 3/2/2/2 & Mylar/$\sim$~0.1  & G10/kapton/honeycomb & 0.32 \\ \hline
  SBS 1     & 3/2/2/2 & Al+kapton         & G10/kapton/honeycomb & 0.345 \\ \hline
  UVA-1m-GEM & 3/2/2/2       & Mylar/$\sim$~0.1 & G10/kapton/Rohacell foam & 0.42 \\ \hline
  FIT-1m-zigzag-GEM & 3/1/2/1 & PCB/3.175        & G10/3.175                       & 3.88 \\ \hline
 FIT-30cm & 3/2/2/2 & PCB/3.175      & G10/2.362                        &3.42 \\ \hline
 FIT-10cm-1&3/2/2/2& Mylar/$\sim$~0.1 & G10/2.362                   & 1.5 \\ \hline
 FIT-10cm-2&3/2/2/2& Honycomb/3.175&G10/2.362                   & 1.48 \\ \hline
 Tracker 3 & 3/2/2/2 & Al+kapton     & G10/kapton/honeycomb   & 0.345 \\ \hline
 Tracker 4 & 3/2/2/2 & Mylar/$\sim$~0.1 & G10/kapton/honeycomb& 0.32 \\ \hline
 Ar/CO$_{2}$ & 88 mm &                 &                                      & $\sim$~0.66 \\ \hline
 Air       & $\sim$~3 m &                  &                                        & $\sim$~1 \\ \hline \hline
 Total Material        &                  &       &                                 &  \textbf{14\% $X_{0}$}    \\
  \hline
\end{tabular}
\label{mytable}
\end{table}

\begin{figure}[h]
  \centering
  \includegraphics[width=10cm]{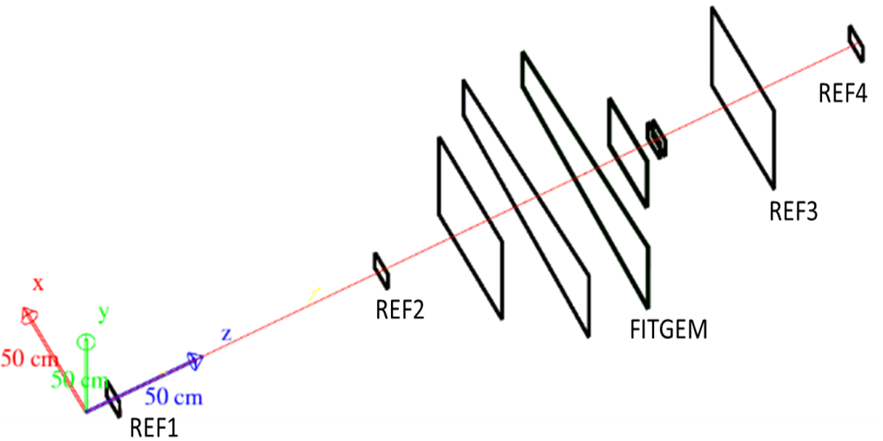}
  \vspace{-0.5cm}
  \caption{The stand-alone Geant4 geometry of all GEM detectors in the FNAL 2013 beam test. The large-area GEM detector
with radial zigzag strips is labelled as FITGEM.}
  \label{fig13}
\end{figure}

A stand-alone Geant4~\cite{G4Reference} simulation was created to study the impact of MCS and to extract the intrinsic detector resolution precisely. The beam test setup as simulated in Geant4 is shown in Fig.~\ref{fig13}. The material distribution from Table~\ref{mytable} is fully implemented in the simulation, while details such as GEM holes and readout strip geometries are not. For the Geant4 physics list, FTFP\_BERT is used, which includes the MCS model based on Lewis theory. Perpendicular point-like beams start 20 mm in front of the first tracker detector (REF1) at position (x,y) = (0 mm, 0 mm) and different beam momenta and particles are simulated.

\begin{figure}[h]
  \begin{center}
     \subfigure{
          \label{fig14:top}%%
          \begin{minipage}[b]{0.5\textwidth}
              \centering
              \includegraphics[width=8cm,height=6cm]{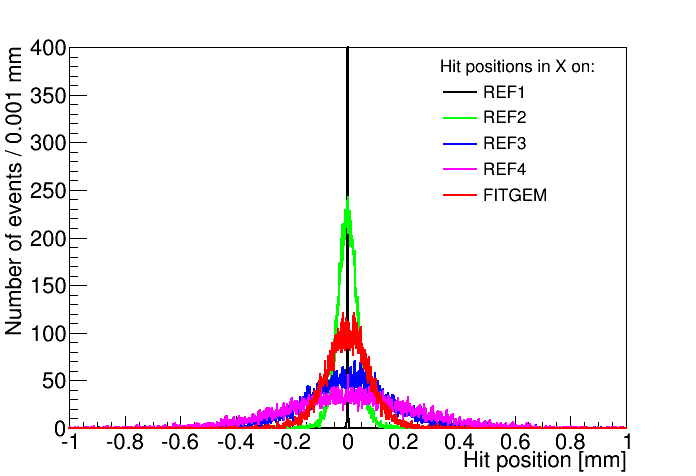}
          \end{minipage}}\newline%
    \subfigure{
          \label{fig14:bottom}%%
          \begin{minipage}[b]{0.5\textwidth}
              \centering
              \includegraphics[width=8cm,height=6cm]{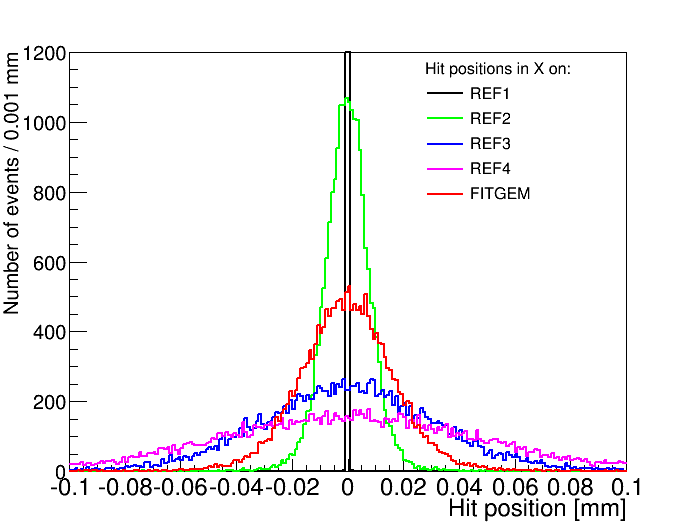}
          \end{minipage}}\newline%
    \vspace{-1cm}
    \caption{Simulated hit positions in the tracker and in the zigzag GEM detector (FITGEM) for 25 GeV/c pions (top) and 120 GeV/c protons (bottom).}
    \label{fig14}
  \end{center}
\end{figure}

\begin{figure}[h]
  \centering
  \includegraphics[width=10cm]{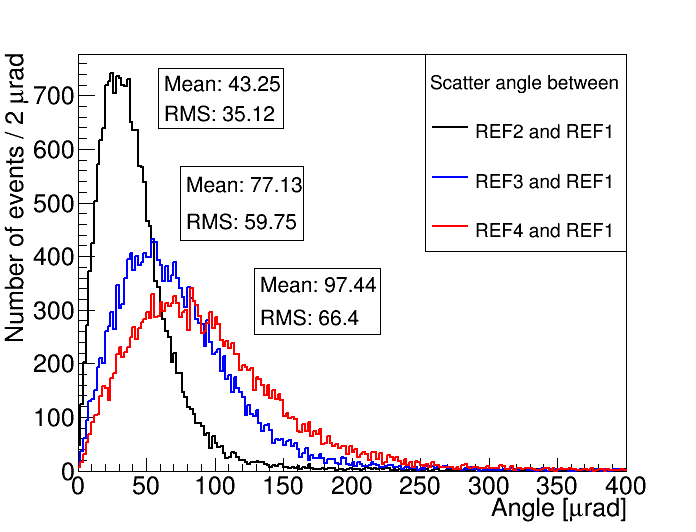}
  \vspace{-0.5cm}
  \caption{Distributions of scattering angles between different reference detectors due to multiple scattering. The mean and rms values in $\mu$rad of each distribution are indicated on the plot.}
  \label{ScatterAngle}
\end{figure}

\begin{figure}[h]
  \centering
  \includegraphics[width=10cm]{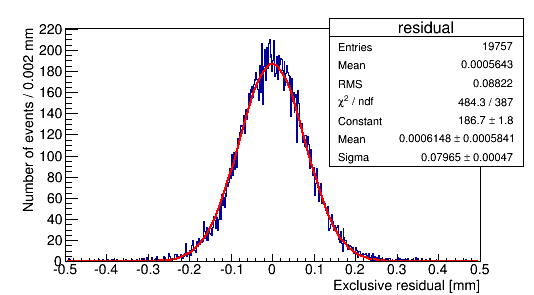}
  \vspace{-0.5cm}
  \caption{Simulated exclusive residual distribution for the zigzag GEM detector for 25~GeV/c pions. All detectors are set to have perfect intrinsic resolution here so that the width of this distribution is a measure of the MCS effect at the position of the zigzag GEM.}
  \label{G4ResFITEICX_Ex}
\end{figure}

We estimate the magnitude of the MCS effect in each detector by running the simulation with perfect intrinsic detector resolutions, but with MCS turned on. The simulated hit positions are histogrammed for each reference tracker detector and for the zigzag GEM detector. In the simulation, the zigzag GEM can be treated as a 2D detector in a Cartesian system. In Fig.~\ref{fig14}, the top plot shows the hit position distributions in the horizontal X direction for 25 GeV/c pions, while the bottom plot is for 120 GeV/c protons. Beams are widened significantly at lower momenta due to MCS and the hit distribution widens as the beam particles move downstream. The scattering angles between detectors can be calculated (Fig.~\ref{ScatterAngle}). The scattering is small at the beginning of the beam and it becomes large at the end of the 3-m long tracking system; the overall mean scattering angle between tracker detectors one and four is about 97~$\mu$rad with a 66~$\mu$rad rms. The exclusive residual width for the zigzag GEM detector due to the MCS effect alone is found to be about 80~$\mu$m in Cartesian coordinates~(Fig.~\ref{G4ResFITEICX_Ex}) when simulating 25~GeV/c pions with all tracker detectors set to perfect intrinsic resolution. Here `exclusive' means that the hit of the zigzag detector is excluded from the track fit.

\subsection{Alignment of reference trackers}

The first step in the resolution measurement is an alignment of the four small tracking detectors that have a Cartesian X-Y strip readout. The trackers are first aligned to each other in Cartesian coordinates. The data sample with the highest statistics is used for this alignment, which was taken with 32~GeV/c mixed hadron beams. Only events with a single cluster observed in each tracker are used for the alignment. Any strip multiplicity is allowed in the clusters including single-strip clusters. 

\begin{figure}[h]
  \centering
  \includegraphics[width=12cm, height=7cm]{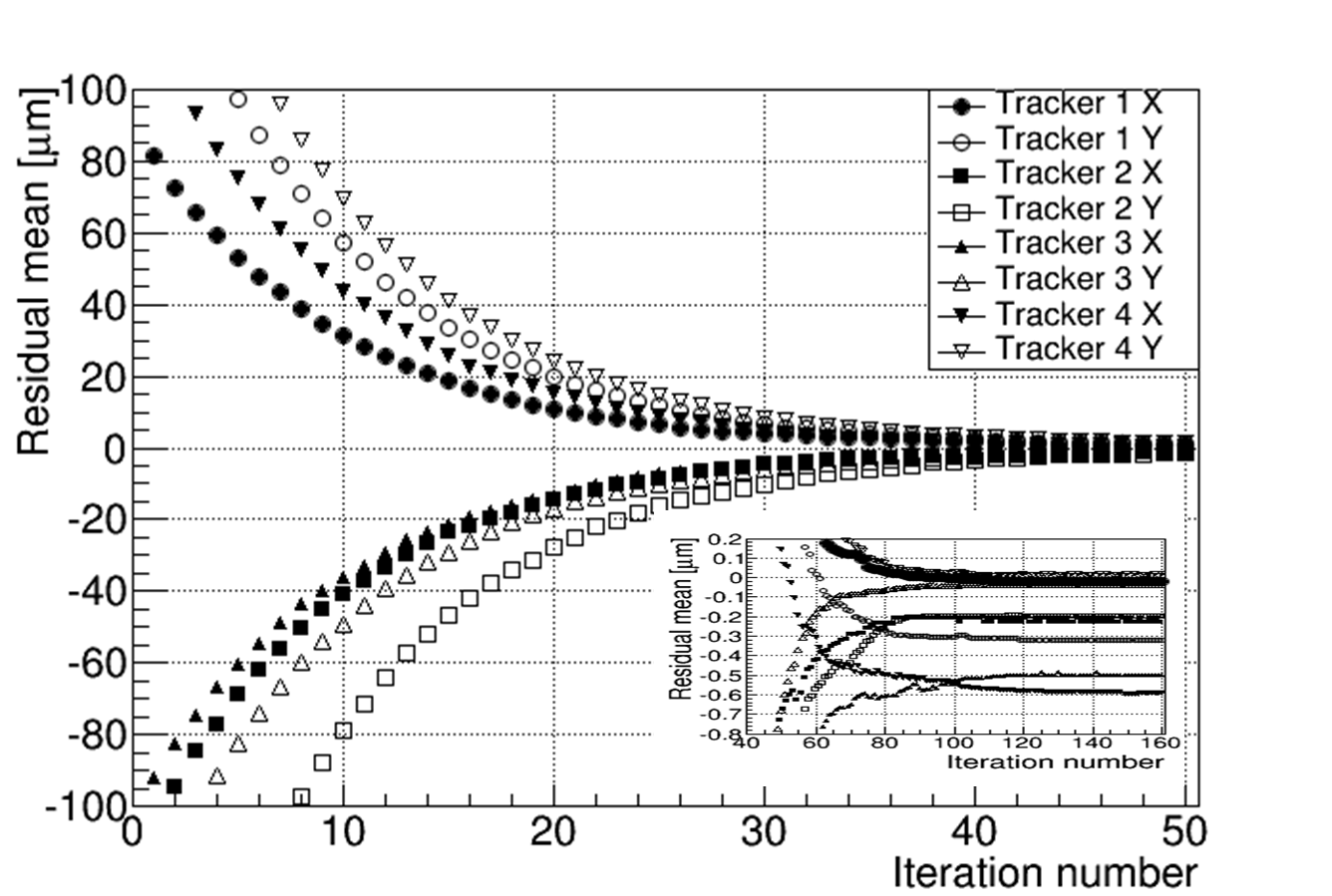}
  \vspace{-0.5cm}
  \caption{Residual means vs.\ iteration number in the first step of aligning the tracker. In this step, the detectors are only shifted in each iteration. The inset is a zoom-in around zero.}
  \label{fig17}
\end{figure}

\begin{figure}[h]
  \centering
  \includegraphics[width=12cm, height=7cm]{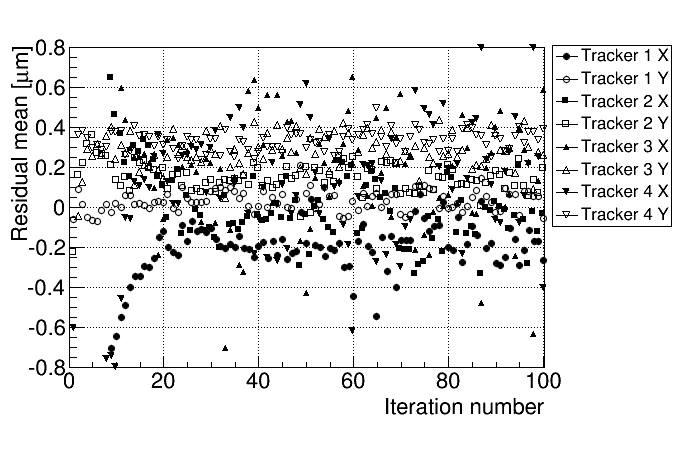}
  \vspace{-0.5cm}
  \caption{Residual means vs.\ iteration number in the second step of aligning the tracker. In this step, detectors are alternately shifted and rotated relative to tracker detector REF1 in each iteration.}
  \label{fig18}
\end{figure}

\begin{figure}[h]
  \begin{center}
     \subfigure{
          \label{fig19:top}%%
          \begin{minipage}[b]{0.5\textwidth}
              \centering
              \includegraphics[width=8cm,height=6cm]{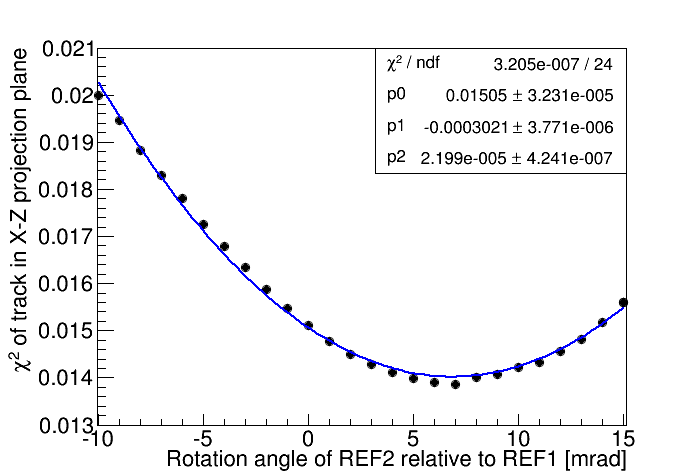}
          \end{minipage}}\newline%
    \subfigure{
          \label{fig19:bottom}%%
          \begin{minipage}[b]{0.5\textwidth}
              \centering
              \includegraphics[width=8cm,height=6cm]{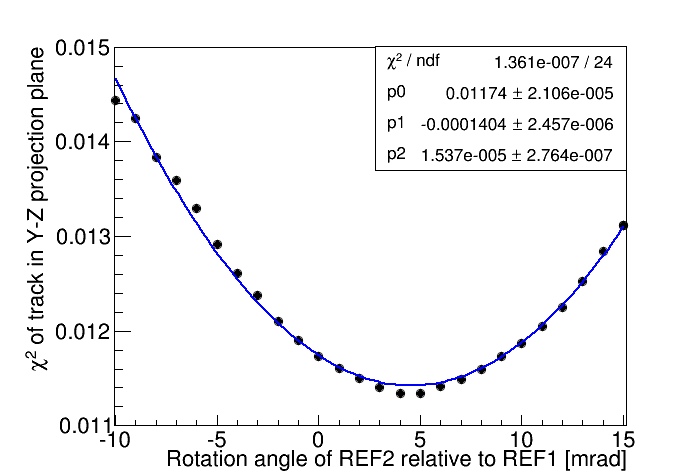}
          \end{minipage}}\newline%
    \vspace{-1cm}
    \caption{Track-$\chi^{2}$ in X-Z (top) and Y-Z (bottom) planes vs.\ rotation angle of tracker detector REF2 relative to tracker detector REF1. This is an example for the third step of aligning the tracker; the final relative rotation angle obtained from averaging the minima of the two parabolic curves is 5.72~mrad.}
    \label{fig19}
  \end{center}
\end{figure}

The first alignment step is to shift each of the four tracking detectors iteratively in the XY-plane to make their origins match each other in that plane. The initial shift parameters are mean values from position distributions in X and Y coordinates. In each iteration, straight lines are fitted to the hits in X and Y. Residuals are histogrammed for each detector and the residual distributions are fitted with a double-Gaussian function. Ten percent of the residual mean value of each detector is taken as the shift parameter in the next iteration to avoid overcorrections. The resulting residual mean values converge quickly towards zero after 40 iterations, as can be seen in Fig.~\ref{fig17}. This provides a first coarse alignment. In a second alignment step, we correct also for relative rotations of the tracking detectors around the beam in the XY-plane. We again fit straight lines to the hits in X and Y and iterate through a succession of offsets and rotations around the beam axis relative to the first tracking detector until the residual means from the track fits are very close to zero and the $\chi^2$ of the track fits are minimized. In each iteration, the detectors are first shifted and then rotated; then new residuals and rotation angles are calculated. After about 20 iterations the residual mean values become flat within $\pm$0.4~$\mu$m around zero (Fig.~\ref{fig18}). From this step, we get eight shift parameters and three relative rotation angles as refined alignment parameters for the tracker system. The final step is to optimize the three rotation angles one-by-one. With the shift parameters and two rotation angles kept fixed, the third rotation angle is changed in a small range around the value from the second step with a 1~mrad step. We find parabolic curves of mean $\chi^{2}$ of tracks vs.\ angle and calculate the optimized rotation angle from the minimum of parabolic fits to those curves (Fig.~\ref{fig19}). The final rotation angle is taken as the average of these angles obtained for X and Y directions.

\subsection{Spatial resolution of reference trackers}\label{TrackErrMethod}

\begin{figure}[h]
  \begin{center}
     \subfigure{
         \label{TrackersRes:top}%%
          \begin{minipage}[b]{0.33\textwidth}
              \centering
              \includegraphics[width=5cm,height=4.5cm]{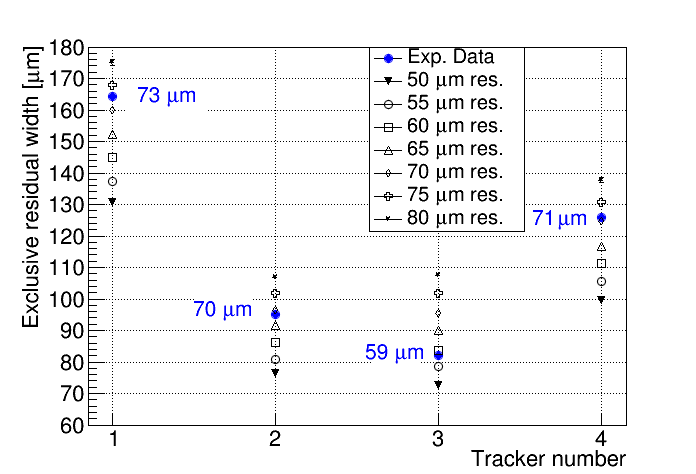}
          \end{minipage}}%
    \subfigure{
          \label{TrackersRes:center}%%
         \begin{minipage}[b]{0.33\textwidth}
            \centering
            \includegraphics[width=5cm,height=4.5cm]{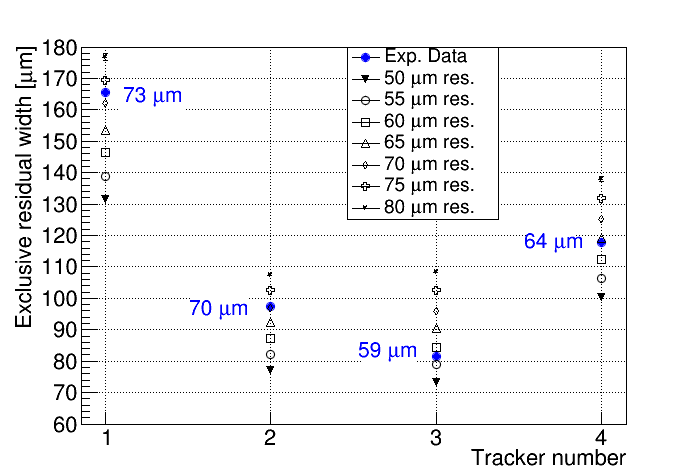}
        \end{minipage}}%
    \subfigure{
          \label{TrackersRes:bottom}%%
         \begin{minipage}[b]{0.33\textwidth}
            \centering
            \includegraphics[width=5cm,height=4.5cm]{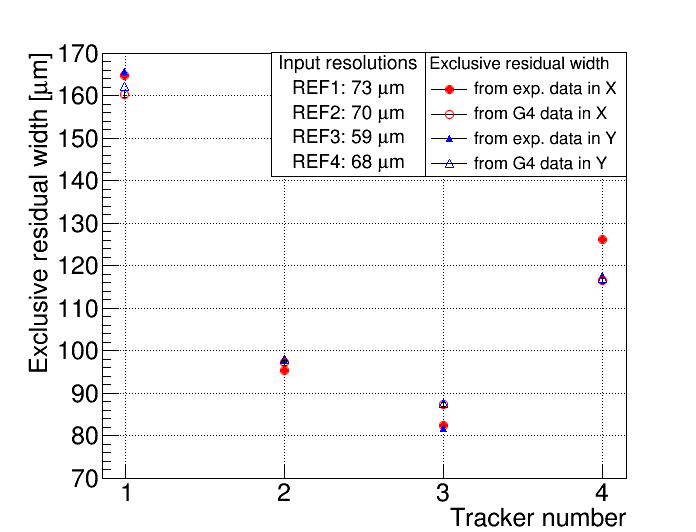}
        \end{minipage}}%
    \vspace{-0.3cm}
    \caption{Exclusive tracker detector residuals for X~(left) and Y~(center) coordinates obtained with common input values for the intrinsic detector resolutions (Gaussian smearing in 5~$\mu$m steps, black points) in the Geant4 simulation with MCS compared with experimental residuals (blue points) in the beam test. The numbers in blue (left and center) are resolutions for the tracker detectors in the beam tests obtained from this comparison with the simulation. When feeding these resolutions (averaged over X and Y) for each tracker detector back into the simulation as inputs for smearing, the resulting simulated residuals are found to be consistent with the experimental data (right)}.
    \label{TrackersRes}
  \end{center}
\end{figure}

The Geant4 simulation of pions traversing the setup allows us to estimate track errors and to measure intrinsic detector resolutions when MCS effects are included. The resolutions of the tracker detectors are studied to ensure that their resolutions are good enough to serve as precision reference detectors. In the first simulation step, the intrinsic detector resolutions are first set to zero (perfect resolution) but with MCS turned on, and then the obtained hit positions are additionally smeared by hand using a Gaussian to simulate the intrinsic detector resolution.
We initially assume that the four tracker detectors have the same resolution and smear the simulated hit data with resolutions from 50-80 $\mu$m in 5 $\mu$m steps. For each smeared resolution, we calculate the exclusive track-hit residuals for each detector and get the residual widths, so we can compare exclusive residual widths from simulation with those observed in experimental data (Fig.~\ref{TrackersRes}). When the residual width matches for a tracker detector, the corresponding input resolution used for the Gaussian smearing is taken as the intrinsic resolution of that detector. We do this in both X and Y coordinates and the average of the two is taken as the final resolution for a tracker detector. The resolutions of the tracking detectors 1-4 are found to be 73, 70, 59, and 68 $\mu$m, respectively. The statistical uncertainties of these resolutions are smaller than 2.5~$\mu$m (half of the step size). These measured resolutions fall into a typical range for standard GEM detectors with 400 $\mu$m strip pitch.

\subsection{Alignment of GEM detector with zigzag strips}

\begin{figure}[h]
  \centering
  \includegraphics[width=6cm, height=4.5cm]{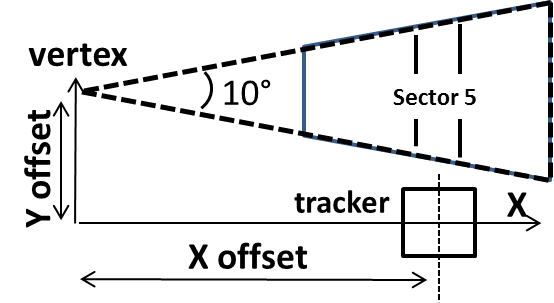}
  \vspace{-0.5cm}
  \caption{Schematic diagram for transformation of the Cartesian tracker coordinates into the natural polar coordinate system of the zigzag GEM detector, which has the vertex of the trapezoid in its origin.}
  \label{fig21}
\end{figure}

Since the zigzag strips in the large GEM detector run in radial direction and measure only the azimuthal $\phi$-coordinate, we need to perform the resolution study for this detector in polar coordinates. The tracker hit positions are transferred to the polar coordinate system that is naturally given by the geometry of the large zigzag detector with the vertex of the trapezoid used as the origin of this polar system (Fig.~\ref{fig21}), and the tracker tracks are then refit in these polar coordinates.

In the polar coordinate system, the $\phi$-coordinate of the center of each zigzag strip in the large trapezoidal detector is simply given by $\phi_{\mathrm{n}} = -0.5\cdot\alpha~+~n \cdot \alpha/(128-1)$, where $\alpha$~=~10$^o$ is the opening angle of the trapezoidal shape and n~=~0,1,$\dots$,127 is the strip number. The $\phi$-position of a hit is initially determined from the barycenter, or centroid, of the cluster using the strip charges as weights. 
    
For transforming the Cartesian tracker coordinates into the polar coordinate system, we need to find the correct X$\mathrm{_{offset}}$ and Y$\mathrm{_{offset}}$ from the new origin to the center of the tracker (Fig.~\ref{fig21}) using the tracks in the $\phi$-coordinate. This has to be done for each test scenario. Note that since the tracker is already internally aligned, the centers of the individual trackers already match, so there is only a single pair (X$\mathrm{_{offset}}$, Y$\mathrm{_{offset}}$) of offsets to be found for each scenario. The idea behind our alignment procedure is that any misalignment will shift the residual means away from zero and will increase the residual widths when fitting tracks in the $\phi$-coordinate with hits in the trapezoidal GEM.
 	
We use a two-fold iteration loop to find the X and Y offsets for each data-taking scenario. First, we keep a fixed X$\mathrm{_{offset}}$ value and step through Y$\mathrm{_{offset}}$ values with a fixed step of 0.2~mm within a reasonable physical range, then change to another X$\mathrm{_{offset}}$ value with a step of 1~mm and vary the Y$\mathrm{_{offset}}$ again until all X$\mathrm{_{offset}}$ values have been covered. For each (X$\mathrm{_{offset}}$, Y$\mathrm{_{offset}}$) pair, tracks are fitted in the $\phi$-coordinate including and excluding the $\phi$-hit in the zigzag detector and the corresponding inclusive and exclusive residuals for the zigzag GEM detector are histogrammed. We record residual means and widths, as well as the mean $\chi^2$ of the track fits in the $\phi$-coordinate. 
    
\begin{figure}[h]
  \begin{center}
     \subfigure{
          \label{fig22:left}%%
          \begin{minipage}[b]{0.33\textwidth}
              \centering
              \includegraphics[width=6cm,height=4.5cm]{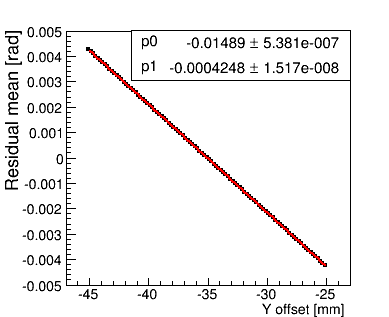}
          \end{minipage}}%
    \subfigure{
          \label{fig22:middle}%%
          \begin{minipage}[b]{0.33\textwidth}
              \centering
              \includegraphics[width=6cm,height=4.5cm]{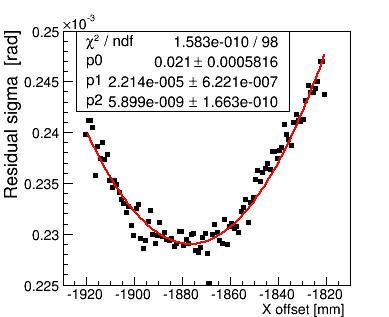}
          \end{minipage}}%
    \subfigure{
          \label{fig22:right}%%
          \begin{minipage}[b]{0.33\textwidth}
              \centering
              \includegraphics[width=6cm,height=4.5cm]{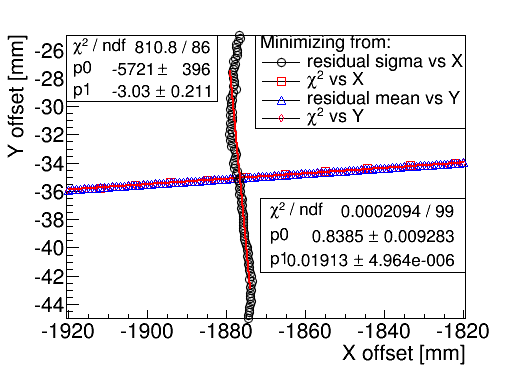}
          \end{minipage}}\newline%
    \vspace{-1cm}
  \caption{Optimization of X and Y offsets for tracker detectors for beam impinging on zigzag GEM in central sector 5. Left: At fixed $\mathrm{X_{offset}=-1876\ mm}$, the residual mean vs.\ $\mathrm{Y_{offset}}$ curve is a line; the best $\mathrm{Y_{offset}}$ at this $\mathrm{X_{offset}}$ point is calculated from requiring that the residual mean be equal to zero: $\mathrm{Y_{offset} =-35.05\ mm}$. Center: At fixed $\mathrm{Y_{offset}=-35\ mm}$, the residual width vs.\ $\mathrm{X_{offset}}$ is a parabola; the best $\mathrm{X_{offset}}$ is calculated from the minimum of this parabola: $\mathrm{X_{offset}=-1876.6\ mm}$. Right: Scatter plots of best $\mathrm{(X_{offset}, Y_{offset})}$ points from residual mean vs.\ $\mathrm{Y_{offset}}$ and residual sigma vs.\ $\mathrm{X_{offset}}$, as well as plots of $\mathrm{\chi^2}$ vs.\ $\mathrm{X_{offset}}$ and $\mathrm{\chi^2}$ vs.\ $\mathrm{Y_{offset}}$, which strongly overlap with the former.}
  \label{fig22}
  \end{center}
\end{figure}

We use a three-step procedure to find the best estimate of the offsets. Fig.~\ref{fig22} shows an example of this procedure for the data recorded in $\eta$-sector five of the zigzag GEM at 3300~V. For a given X$\mathrm{_{offset}}$ value we record that Y$\mathrm{_{offset}}$ value which produces a well-centered $\phi$-residual for the zigzag detector with a mean of zero (Fig.~\ref{fig22}, left). Then, for a given Y$\mathrm{_{offset}}$, we plot the residual widths as function of X$\mathrm{_{offset}}$ and find the X$\mathrm{_{offset}}$ value that minimizes the residual width using a parabolic fit (Fig.~\ref{fig22}, center). This yields two sets of (X$\mathrm{_{offset}}$, Y$\mathrm{_{offset}}$) pairs that are plotted as two curves. The (X$\mathrm{_{offset}}$, Y$\mathrm{_{offset}}$) values for which those two curves intersect is taken as the best estimate of the alignment offsets (Fig.~\ref{fig22}, right). Best (X$\mathrm{_{offset}}$, Y$\mathrm{_{offset}}$) pairs based on track-$\chi^2$ are also plotted and confirm the results.  
	
\begin{figure}[h]
  \begin{center}
     \subfigure{
          \label{fig23:top}%%
          \begin{minipage}[b]{0.5\textwidth}
              \centering
              \includegraphics[width=6cm,height=4cm]{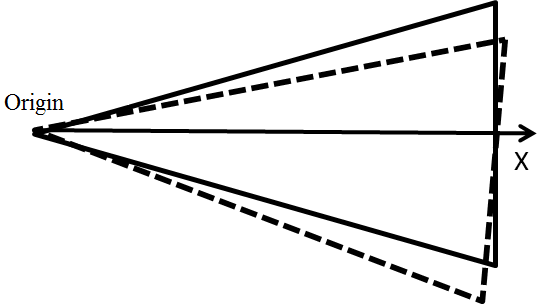}
          \end{minipage}}\newline%
    \subfigure{
          \label{fig23:bottom}%%
          \begin{minipage}[b]{0.5\textwidth}
              \centering
              \includegraphics[width=6.5cm,height=5cm]{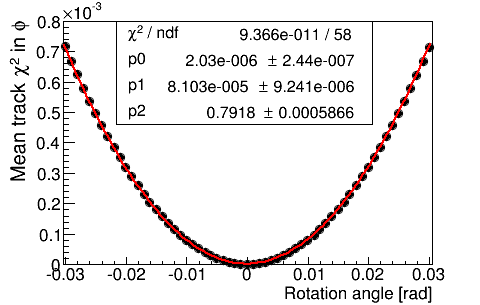}
          \end{minipage}}\newline%
    \vspace{-1cm}
    \caption{Top: Schematic diagram for slight relative rotation of the zigzag GEM detector. Bottom: Mean $\chi^{2}$ of tracks in $\phi$ coordinate vs.\ rotation angles for the same data as in Fig.~\ref{fig22}; from the minimum of the parabola the global rotation angle of the zigzag GEM is found to be $-51~\mu$rad.}
    \label{fig23}
  \end{center}
\end{figure}

Finally, it is not guaranteed that tracker and zigzag detector are installed without a relative rotation in the X-Y plane (Fig.~\ref{fig23}, top). Since aligning the zigzag GEM detector also requires matching the X direction to the trackers so that all calculations of $\phi$ are correct, the rotation angle of the zigzag GEM detector needs to be checked. The mean $\chi^2$ of tracks in $\phi$ vs.\ rotation angles is a parabola and the angle can be calculated from the minimum; it turns out to be very close to zero (Fig.~\ref{fig23}, bottom).

\subsection{Corrections for non-linear strip response of the GEM detector with zigzag strips}

Before proceeding to the measurement of the angular resolution for the zigzag GEM detector, we correct its hit data. The reason to do so is that there is a non-linear response of the zigzag strips when the hit positions are calculated from the centroids of clusters. Specifically, we use track information to apply the non-linear corrections to the hit positions. For each  cluster with a strip multiplicity N~$>$~1, we define a quantity $\eta = \mathrm{s_c-s_{max}}$, where $\mathrm{s_c} = \Sigma_{i=1}^{n}\mathrm{q_i\cdot s_i} / \Sigma_{i=1}^{n} \mathrm{q_i}$ is the centroid position of the cluster in terms of strip number; $\mathrm{s_i}$ and $\mathrm{q_i}$ are the number and the induced charge (in ADC counts) for the i$^\mathrm{th}$ strip in the cluster and $\mathrm{s_{max}}$ is the number of the strip in the cluster on which the maximum charge is induced. The quantity $\eta$ is then a measure of the difference between centroid position and the maximum-charge strip independent of the absolute strip number. We treat different strip multiplicities separately because $\eta$ has different characteristics for odd and even strip multiplicities. For even multiplicities, $s_c$ tends to be between two strips, whereas for odd multiplicities it is close to the center of the strip with maximum charge. In practice, we only need to correct clusters with strip multiplicities N~=~2 and N~=~3 using corresponding quantities $\eta_2$ and $\eta_3$. Corrections for N~$>$~3 are not necessary since there are very few of such clusters (see also Fig.~\ref{fig8}).

\begin{figure}[h]
  \begin{center}
     \subfigure{
          \label{fig24:top}%%
          \begin{minipage}[b]{0.5\textwidth}
              \centering
              \includegraphics[width=8cm,height=6cm]{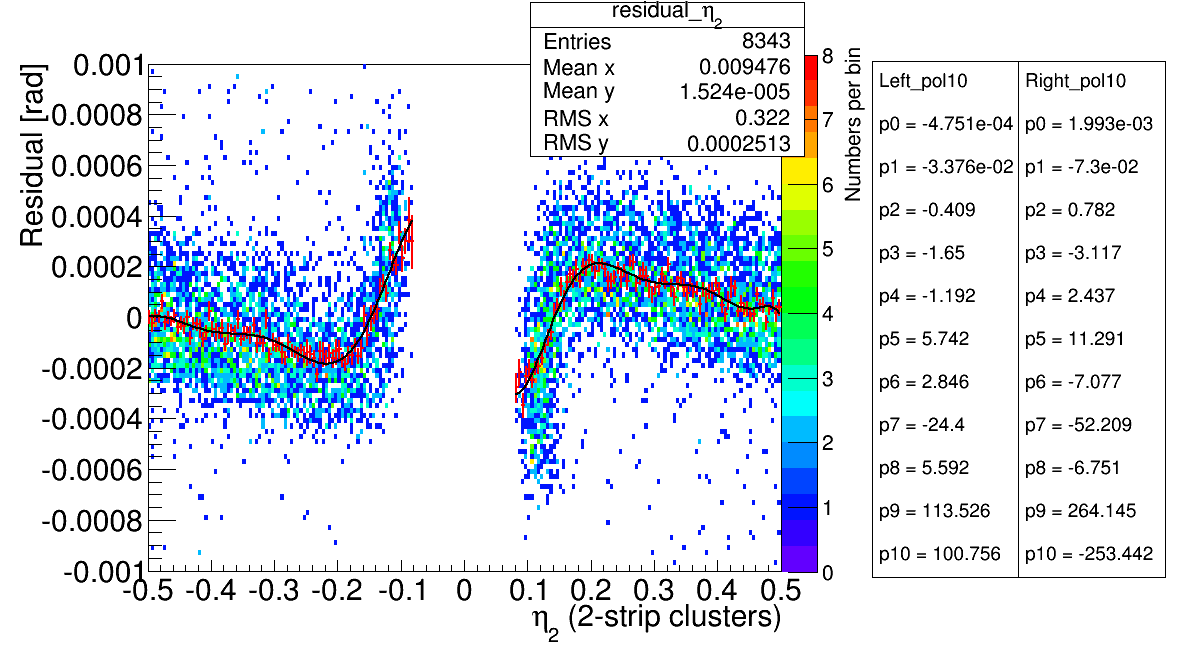}
          \end{minipage}}\newline%
    \subfigure{
          \label{fig24:bottom}%%
          \begin{minipage}[b]{0.5\textwidth}
              \centering
              \includegraphics[width=8cm,height=6cm]{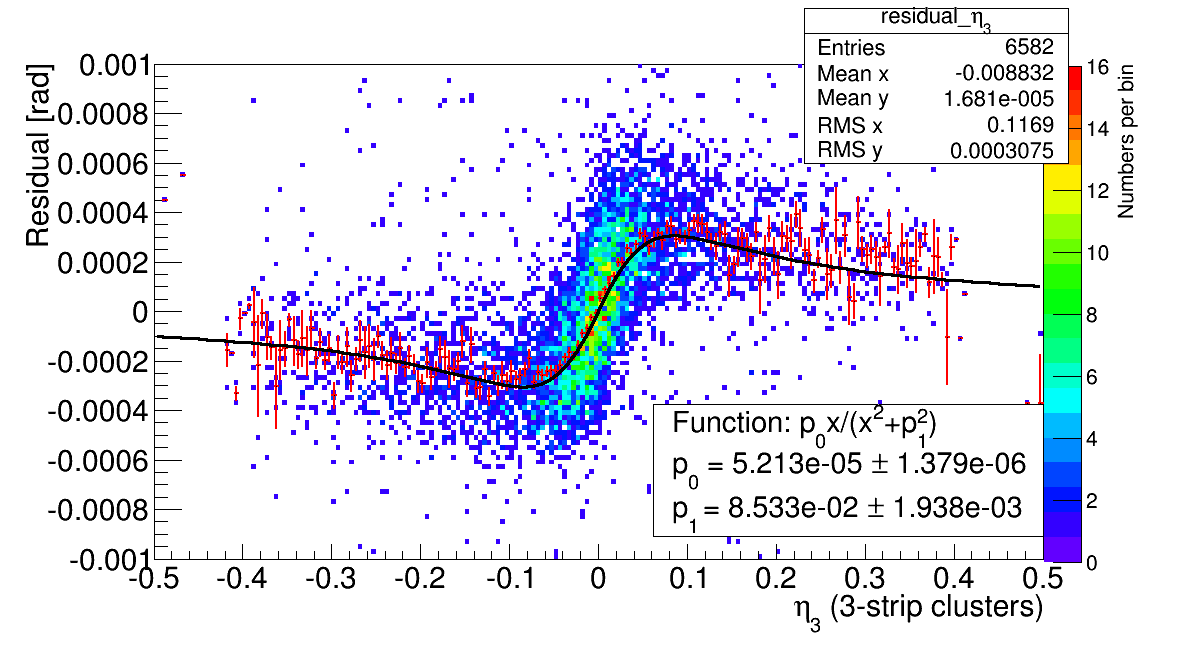}
          \end{minipage}}\newline%
    \vspace{-1cm}
    \caption{Top: Exclusive residuals of 2-strip clusters in zigzag GEM vs.\ $\mathrm{\eta_2}$ fitted with 10-degree polynomials. Bottom: Exclusive residuals of 3-strip clusters in zigzag GEM vs.\ $\mathrm{\eta_3}$ fitted with a serpentine function.}
    \label{fig24}
  \end{center}
\end{figure}

\begin{figure}[h]
  \begin{center}
     \subfigure{
          \label{fig25:top}%%
          \begin{minipage}[b]{0.5\textwidth}
              \centering
              \includegraphics[width=8cm,height=6cm]{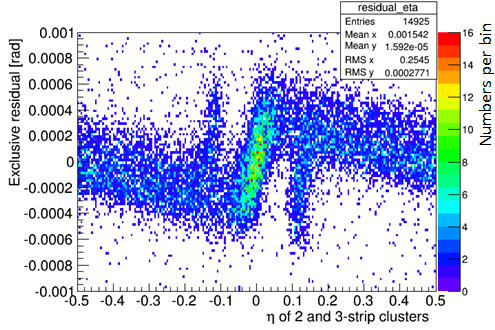}
          \end{minipage}}\newline%
    \subfigure{
          \label{fig25:bottom}%%
          \begin{minipage}[b]{0.5\textwidth}
              \centering
              \includegraphics[width=8cm,height=6cm]{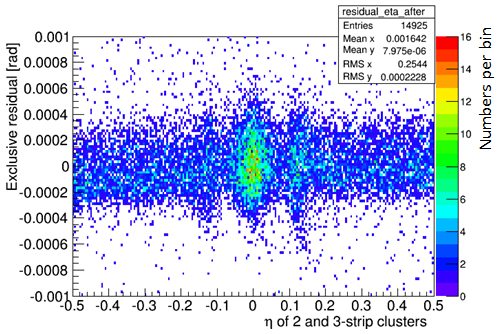}
          \end{minipage}}\newline%
    \vspace{-1cm}
    \caption{Exclusive residuals of zigzag GEM vs.\ $\mathrm{\eta}$ for strip multiplicities N=2 and N=3 clusters before (top) and after (bottom) correction.}
    \label{fig25}
  \end{center}
\end{figure}

Using reference tracks given by the tracker, exclusive residuals of hits in the zigzag GEM are then plotted vs.\ $\eta$. We observe that the resulting distribution is not flat, which indicates a non-linear response of the zigzag strips to the true hit position. As shown in Fig.~\ref{fig24}, exclusive residuals are plotted vs.\ $\eta_2$ and $\eta_3$ and their profiles are fitted separately to appropriate fit functions. A 10$^{\mathrm{th}}$-degree polynomial produces a good fit for the $\eta_2$ distribution (Fig.~\ref{fig24}, top), while a serpentine function is used for the $\eta_3$ distribution (Fig.~\ref{fig24}, bottom). We then globally correct the orginal hit positions by subtracting the $\eta$-dependent offsets obtained from the fit functions from the original hit positions. Fig.~\ref{fig25} shows that the exclusive residual vs.\ $\eta$ plot is much flatter after this correction. Consequently, the overall residual distribution becomes significantly narrower and the spatial resolution is improved when the corrected hit positions are used. Note that the tracker detectors do not need to be corrected since their resolutions are already good enough as they have straight strips with much smaller pitch and consequently much more linear responses. For the position scan data, we apply individual correction functions for each sector since the strip width changes along the radius. For the HV scan data, which were taken in a single sector, the correction functions are obtained from all HV scan data combined together and applied equally to different voltage points. We have checked that the difference is very small if we find correction functions using only data taken at one voltage point.

\subsection{Angular resolution measurement for GEM detector with zigzag strips}
\begin{figure}[h]
  \centering
  \includegraphics[width=10cm]{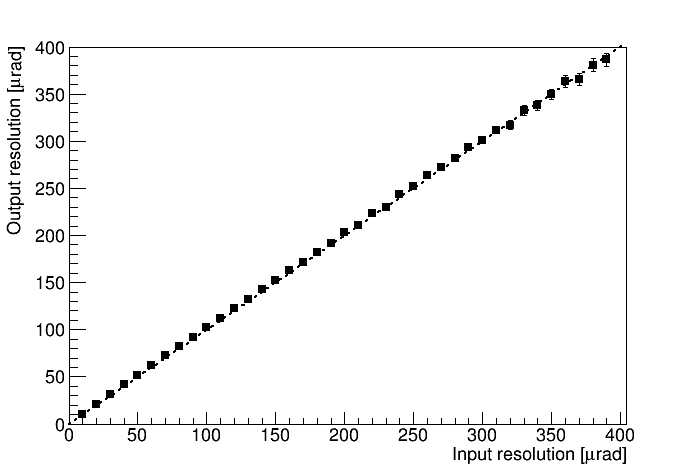}
  \vspace{-0.5cm}
  \caption{Intrinsic angular resolutions calculated using track error estimate for the large GEM detectors vs.\ the smeared input resolutions in simulation, showing very good agreement. Errors are smaller than marker size for most of the points.}
  \label{figG4SimFITEIC}
\end{figure}

In polar coordinates, we smear the simulated tracker hits with the realistic intrinsic tracker resolutions obtained above and with MCS fully taken into account. We fit a linear track to the azimuthal coordinates of the four smeared hits from the tracker detectors and simulate an unsmeared hit for the zigzag GEM detector. The resulting width of the exclusive residual $\phi_{\mathrm{interpol. track}} - \phi_{\mathrm{unsmear.zigzag}}$ distribution for the zigzag GEM is then a measure of the interpolated-track error (IE) at the position of the zigzag GEM with MCS taken into account. For the experimental data, we also fit linear tracks to the $\phi$-coordinate and calculate the exclusive residual width (ER) for the zigzag GEM detector. The intrinsic resolution $\sigma$ for the zigzag GEM can then be calculated by subtracting the two quantities in quadrature: $\sigma = \mathrm{\sqrt{ER^2-IE^2}}$. We check that this method gives accurate results using the simulation by also smearing the resolution of the probed zigzag GEM detector from 10 to 390 $\mu$rad in 10 $\mu$rad steps and by then applying the same method. We find almost perfect agreement between input resolution from Gaussian smearing and calculated intrinsic resolution $\sigma$ over the full range (Fig.~\ref{figG4SimFITEIC}).

\begin{figure}[h]
  \begin{center}
     \subfigure{
          \label{fig27:2}%%
          \begin{minipage}[b]{0.5\textwidth}
              \centering
              \includegraphics[width=8cm,height=6cm]{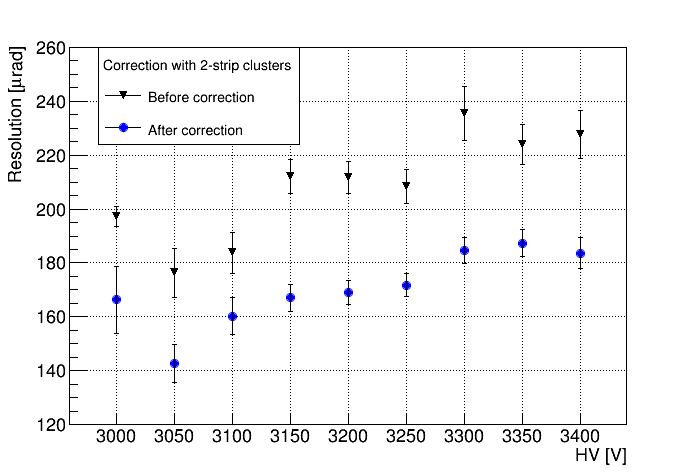}
          \end{minipage}}%
    \subfigure{
          \label{fig27:3}%%
          \begin{minipage}[b]{0.5\textwidth}
              \centering
              \includegraphics[width=8cm,height=6cm]{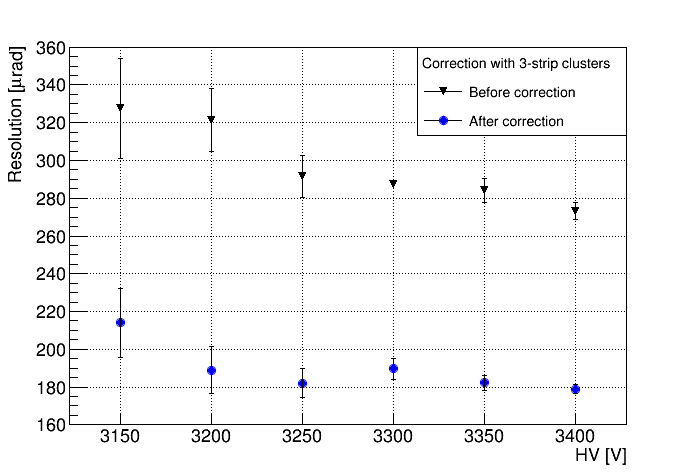}
          \end{minipage}}\newline%
    \subfigure{
          \label{fig27:23}%%
          \begin{minipage}[b]{0.5\textwidth}
              \centering
              \includegraphics[width=8cm,height=6cm]{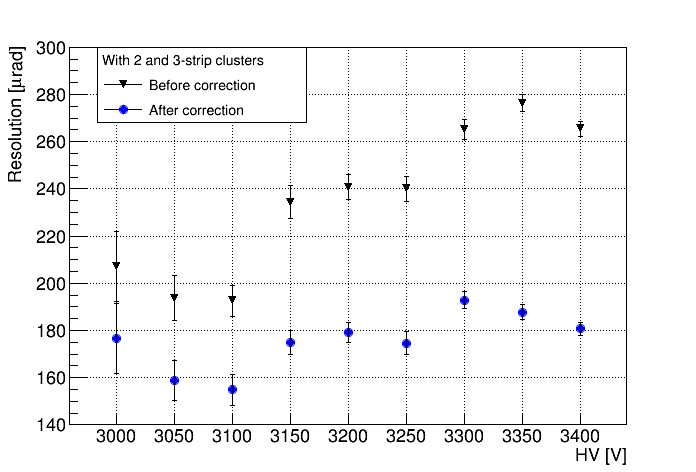}
          \end{minipage}}%
    \subfigure{
          \label{fig27:all}%%
          \begin{minipage}[b]{0.5\textwidth}
              \centering
              \includegraphics[width=8cm,height=6cm]{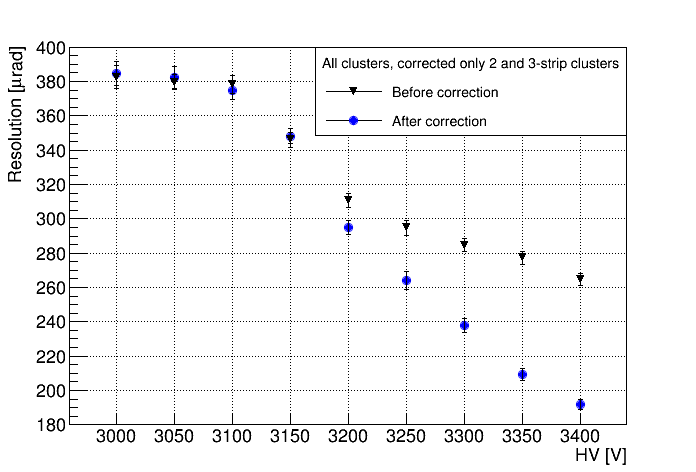}
          \end{minipage}}\newline%
    \vspace{-1cm}
  \caption{Angular resolutions vs.\ $\mathrm{V_{drift}}$ measured in central sector 5 of the zigzag GEM before (black) and after (blue) non-linear response corrections. The plots show results for different strip multiplicities of the clusters.}
  \label{fig27}
  \end{center}
\end{figure}

\begin{figure}[h]
  \begin{center}
     \subfigure{
          \label{fig28:top}%%
          \begin{minipage}[b]{0.5\textwidth}
              \centering
              \includegraphics[width=8cm,height=6cm]{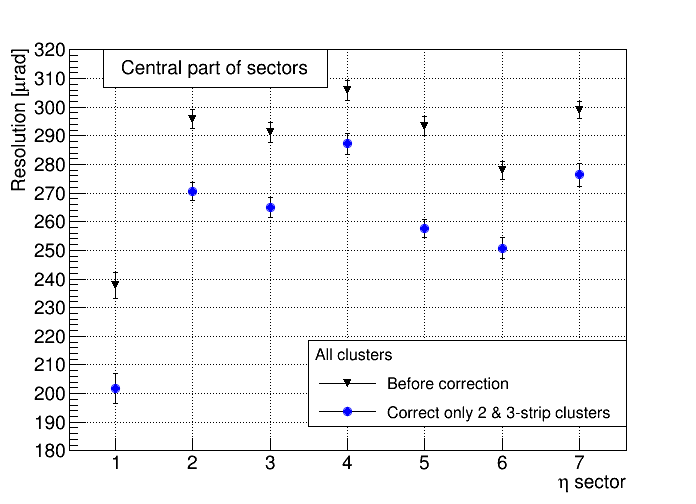}
          \end{minipage}}\newline%
    \subfigure{
          \label{fig28:bottom}%%
          \begin{minipage}[b]{0.5\textwidth}
              \centering
              \includegraphics[width=8cm,height=6cm]{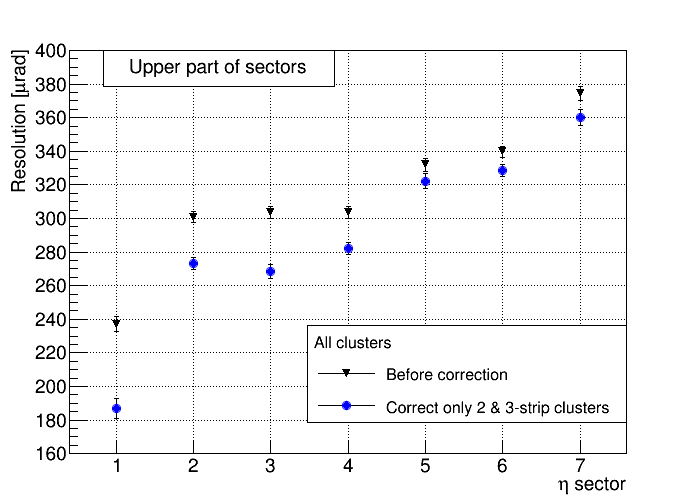}
          \end{minipage}}\newline%
    \vspace{-1cm}
    \caption{Angular resolutions measured at $\mathrm{V_{drift}=3200~V}$ in different positions of the zigzag GEM  before and after corrections using all strip multiplicities. The top (bottom) plot is for positions in the central (upper, i.e.\ $\sim$60 mm above central) part of each sector.}
    \label{fig28}
  \end{center}
\end{figure}

Fig.~\ref{fig27} shows the resulting measured intrinsic angular resolutions for the zigzag GEM detector in central sector five at different $\mathrm{V_{drift}}$ values. The resolution is around 180~$\mu$rad on the efficiency plateau for 2-strip and 3-strip clusters when the non-linear responses of zigzag strips are corrected. In the bottom right plot, we see that the overall angular resolution for the zigzag GEM including single-strip clusters, which cannot be corrected, is $\sigma =$~(193~$\pm$~3)~$\mu$rad at highest tested voltage. This value corresponds to 14\% of the angular pitch of the radial strips. The resolutions are also measured for different positions on the zigzag GEM operated at 3200~V (Fig.~\ref{fig28}).

\section{Summary and Conclusion}
The large-area zigzag GEM detector performed very well in the Fermilab beam test. The $\sim$5.5\% average crosstalk between the zigzag strips is small and does not show significant impact on signal collection. The detection efficiency is above 98\% for charged particles as expected for a GEM detector. The zigzag GEM detector achieves an angular spatial resolution of 193~$\mu$rad or $\sim$14\% of the angular strip pitch after the centroid positions of clusters are corrected for non-linear strip response.

In conclusion, the zigzag readout design is considerably more cost-effective for large-area GEM detectors than conventional straight strip readout structures while preserving good performance. Consequently, GEM detectors with zigzag readout strips are a viable option for application as affordable tracking detectors at future colliders such as the Electron-Ion Collider~(EIC).

\section*{Acknowledgments}
This work is supported by Brookhaven National Laboratory under the EIC eRD-6 consortium. We acknowledge the Fermilab Test Beam Facility staff for their great assistance during the beam test. We are grateful to Alexander Kiselev (Brookhaven National Lab) and Kondo Gnanvo (University of Virginia) for helpful discussions and suggestions on estimating track errors for the resolution studies. Finally, we also would like to thank Bob Azmoun (Brookhaven National Lab) for providing the double-Gaussian fit script and Thomas Hemmick (Stony Brook University) for the discussion on implementing the non-linear hit corrections.

\end{document}